\documentclass{aa}  

\usepackage{graphicx}
\usepackage{txfonts}
\begin{document}

   \title{Association of IceCube neutrinos with radio sources observed at Owens Valley and Mets\"ahovi Radio Observatories\thanks{Flux density data of the associated sources at 15 and 37\,GHz are only available in electronic form at the CDS via anonymous ftp to cdsarc.u-strasbg.fr (130.79.128.5) or via http://cdsweb.u-strasbg.fr/cgi-bin/qcat?J/A+A/}}

   \author{T.~Hovatta
          \inst{1,2}
          \and
          E.~Lindfors\inst{1} 
          \and 
          S.~Kiehlmann\inst{3,4}
          \and
          W.~Max-Moerbeck\inst{5}
          \and
          M.~Hodges\inst{6}
          \and
          I.~Liodakis\inst{1} 
          \and
          A.~L\"ahteem\"aki\inst{2,7} 
          \and
          T.~J.~Pearson\inst{6}
          \and
          A.~C.~S. Readhead\inst{6}
          \and
          R.~A.~Reeves\inst{8}
          \and
          S.~Suutarinen\inst{2} 
          \and
          J.~Tammi\inst{2}
          \and
          M.~Tornikoski\inst{2}
          }
   \institute{Finnish Centre for Astronomy with ESO, FINCA, University of Turku, Finland\\
              \email{talvikki.hovatta@utu.fi, elilin@utu.fi}
         \and
             Aalto University Mets\"ahovi Radio Observatory, Mets\"ahovintie 114, FI-02540 Kylm\"al\"a, Finland
        \and
        Institute of Astrophysics, Foundation for Research and Technology-Hellas, GR-71110 Heraklion, Greece
        \and
    Department of Physics, Univ. of Crete, GR-70013 Heraklion, Greece
    \and
    Departamento de Astronom\'ia, Universidad de Chile, Camino El Observatorio 1515, Las Condes, Santiago, Chile
    \and
    Owens Valley Radio Observatory, California Institute of Technology, Pasadena, CA 91125, USA
    \and
    Aalto University Department of Electronics and Nanoengineering, PO Box 15500, 00076 Aalto, Finland
    \and
    CePIA, Astronomy Department, Universidad de Concepci\'on, Casilla 160-C, Concepci\'on, Chile
    \\}

   \date{Received 21 September 2020; accepted 4 April 2021}

\titlerunning{OVRO and Mets\"ahovi radio sources associated with IceCube neutrinos}
\authorrunning{Hovatta, Lindfors et al.}
 
  \abstract
   {Identifying the most likely sources for high-energy neutrino emission has been one of the main topics in high-energy astrophysics ever since the first observation of high-energy neutrinos by the IceCube Neutrino Observatory. Active galactic nuclei with relativistic jets, also known as blazars, have been considered to be one of the main candidates because of their ability to accelerate particles to high energies.}
   {We study the connection between radio emission and IceCube neutrino events using data from the Owens Valley Radio Observatory (OVRO) and Mets\"ahovi Radio Observatory blazar monitoring programs.}
   {We identify sources in our radio monitoring sample that are positionally consistent with IceCube high-energy neutrino events. We estimate their mean flux density and variability amplitudes around the neutrino arrival time, and compare these with values from random samples to establish the significance of our results.}
   {We find radio source associations within our samples with 15 high-energy neutrino events detected by IceCube. Nearly half of the associated sources are not detected in the $\gamma$-ray energies, but their radio variability properties and Doppler boosting factors are similar to the $\gamma$-ray detected objects in our sample, meaning that they could still be potential neutrino emitters. We find that the number of strongly flaring objects in our statistically complete OVRO samples is unlikely to be a random coincidence (at $2\sigma$ level).}
   {Based on our results, we conclude that although it is clear that not all neutrino events are associated with strong radio flaring blazars, observations of large-amplitude radio flares in a blazar at the same time as a neutrino event are unlikely to be a random coincidence.}


   \maketitle
%

\section{Introduction}

Blazars, active galactic nuclei (AGNs) with the relativistic jet pointing close to our line of sight, are among the main candidate sources of astrophysical high energy neutrinos \citep[see e.g.,][]{eichler79, berezinskii81,1987ApJ...320L..81S,stecker91}. The first $\gamma$-ray detection of blazar 3C~279 \citep{1992ApJ...385L...1H} triggered the first estimates of neutrino fluxes from blazars, where the $\gamma$-ray emission was assumed to be of hadronic origin \citep{1992A&A...260L...1M}, and many followed \citep[e.g.,][]{2001ICRC....3.1153M}. 

The first observation of high-energy neutrinos by the IceCube Neutrino Observatory \citep{2013PhRvL.111b1103A,2013Sci...342E...1I} immediately triggered searches for counterpart sources \citep[e.g.,][]{2014MNRAS.443..474P}, but no significant association with  $\gamma$-ray blazars was found \citep{2015MNRAS.451..323B, 2017ApJ...835...45A}. In September 2017,  an  event  designated  IceCube-170922A was observed. The best-fit  reconstructed  direction  was at  0.1$^\circ$ from the  sky  position  of  the  BL  Lac  object TXS~0506+056, which was flaring in the $\gamma$-ray and very high-energy $\gamma$-ray bands \citep{2017ATel10791....1T,2017ATel10817....1M}. The positional and temporal coincidences of the events were high, which suggests that the neutrino originated from this flaring blazar \citep{2018Sci...361..147I}. Several papers  modeling the event showed that sufficient neutrino flux to explain the detection of the event could be produced within the source, even if the leptonic processes dominated the electromagnetic emission in $\gamma$-rays with only subdominant contribution from cascade emission in very high-energy $\gamma$-rays \citep[see e.g.,][]{2018ApJ...863L..10A, 2020ApJ...891..115P,murase18, keivani18}.

While the association of neutrino events and $\gamma$-ray emission has been receiving significant attention, the connection has also proven to be very complicated. The regions that would host sufficiently dense external photon fields to produce high neutrino fluxes (such as those observed from TXS~0506+056 in 2014-2015 \citep{2018Sci...361.1378I}) would also imply strong internal absorption of $\gamma$-rays \citep[see e.g.,][]{2019Galax...7...20B,2019ApJ...881...46R}, and therefore no temporal coincidence would be expected. Indeed, no $\gamma$-ray flare was detected from TXS~0506+056 during the 2014-2015 neutrino burst \citep{2018Sci...361.1378I}.  Based on a sample of three sources, \cite{kun2020} also suggested that the $\gamma$-ray emission may be suppressed, as they see a local decrease in the $\gamma$-ray flux during the time of the neutrino arrival.

It is also possible that not all blazar classes are equally "good" neutrino candidates \citep[e.g.,][]{PhysRevD.66.123003}.  Blazars are classified as flat-spectrum radio quasars (FSRQs) and BL Lacertae objects (BL Lacs) based on optical spectra. They can also be classified based 
on their synchrotron peak frequency. The low synchrotron peak sources (LSPs) have $\nu_{\mathrm S}<10^{14}$ Hz, the intermediate synchrotron peak sources (ISPs) $10^{14}\,\,{\rm Hz}<\nu_{\mathrm S}<10^{15}$ Hz, and high synchrotron peak sources  (HSPs) $\nu_{\mathrm S}>10^{15}$ Hz \citep{abdo10b}. In FSRQs, the synchrotron peak is typically in the infrared regime (i.e., they are LSPs) while for BL Lacs it can be anywhere between infrared and hard X-rays (LBLs, IBLs, and HBLs).
\citet{2016MNRAS.457.3582P} found the most significant connection with HSP sources. The most recent revision (with more data) suggested a connection with ISP and HSP sources \citep{2020MNRAS.497..865G}. On the other hand, \cite{2019ICRC...36..916H} found the most significant connection (though only at $1.9\sigma$ level) with low and intermediate synchrotron peaked BL Lac objects (LBL/IBLs) and the connection to high synchrotron peaked sources had a much lower significance (only $0.5\sigma$).  In earlier studies \citep[e.g.,][]{atoyan03,murase14,dermer14}, it was found that FSRQs would be the best candidates for the diffuse PeV neutrino background due to their broad line region photon fields.

In the present paper we investigate the connection between radio bright AGNs and astrophysical high-energy neutrinos. 
In AGN jets, the radio emission is synchrotron self-absorbed close to the central engine and therefore the emission we observe originates several parsecs downstream from the jet \citep{1978Natur.276..768R,1979ApJ...231..299R,blandford79}. However, radio emission is a good proxy for the general jet activity. For example, radio flux densities are often seen to increase before the flaring at higher energy bands begins \citep[e.g.,][]{lahteenmaki03, 2011A&A...532A.146L}.
This could for example be due to a long-term increase in the total jet power. The increasing jet power would trigger current-driven kink instabilities and lead to strong magnetic dissipation and efficient particle acceleration \citep{2017Galax...5...64N}, which is  also crucial in neutrino production. Furthermore, even though the fourth catalog of AGNs (4LAC) detected by {\it Fermi}-LAT $\gamma$-ray satellite includes 3207 objects \citep{2019arXiv190510771T}, not all blazars are $\gamma$-ray detected. For example, only 444 of the 1157 sources (38\%) in the Candidate Gamma-Ray Blazar Survey (CGRaBS) sample \citep{healey08} currently monitored at Owens Valley Radio Observatory (OVRO) are included in the 4LAC catalog.

There have been some reports on the connection between radio-selected blazars inside IceCube neutrino positional error circles. \cite{krauss14} studied the six most radio-bright blazars within the error regions of PeV IceCube neutrino events and found that the six sources alone were sufficient to explain the PeV neutrino flux.  \cite{2016NatPh..12..807K} discovered PKS B1424$-$418 flaring in the radio and $\gamma$-ray bands in temporal and positional coincidence with the IceCube neutrino IC35, but this neutrino event had a large positional error circle of $15.6^\circ$. 
The first systematic search for radio sources was performed only recently by \citet{2020ApJ...894..101P}, who investigated the association of neutrinos with radio-bright AGNs using the VLBI-based radio fundamental catalog (RFC){\footnote{
\url{http://astrogeo.org/vlbi/solutions/rfc_2019c/}}} and RATAN-600 \citep{1979S&T....57..324K} data.  \citet{2020ApJ...894..101P} found that AGNs positionally associated with high-energy IceCube events typically have stronger parsec-scale cores compared to the rest of the sample. Moreover, they found an average increase of radio emission at frequencies above 10 GHz around neutrino arrival times for several AGNs on the basis of RATAN-600 monitoring.  The connection was strongest for their highest frequency data at 22\,GHz.  \cite{plavin20b} further analyzed the IceCube point-source likelihood map, including lower energy events, with a comparison to the RFC catalog and found a 3$\sigma$ significance for the correlation between them.

In this paper we investigate the connection between radio emission and neutrinos using the high-cadence long-term monitoring data of the Owens Valley Radio Observatory \citep{2011ApJS..194...29R} and Mets\"ahovi Radio Observatory \citep{2004A&A...427..769T}. Our paper is organized as follows. In Sect.~\ref{data}, we describe the selection of the IceCube neutrino events and our radio samples. Our analysis methods and results are given in Sect.~\ref{analyses} and we discuss them in the context of other recent studies in Sect.~\ref{discussion}. Our conclusions are drawn in Sect.~\ref{conclusions}.

\section{Data}\label{data}

\subsection{Neutrino events}

High-energy neutrinos can interact with ice creating high-energy leptons via charged current interactions. For association with astrophysical sources, the track-like events created by high-energy muons are particularly important, as they have the smallest error circle around the arrival direction.  The angular error can be as small as $0.4^\circ$ \citep{2017APh....92...30A} with typical values on the order of $1^\circ$.

While most of the events are associated with the atmospheric background, some high-energy events have a good chance of being of astrophysical origin. In general, events with energy $>200$\,TeV have likelihood $>50\%$ of being of astrophysical origin \citep{2019arXiv190711699I}. There are also indications that the observed neutrino spectrum hardens above 200\,TeV \citep{2017arXiv171001191I}, while below 200\,TeV there could also be a contribution from the Galactic component \citep{2016ApJ...826..185P}. 

The IceCube Collaboration has published lists of high-energy neutrino candidates \citep{2014PhRvL.113j1101A,2015arXiv151005223T,2016ApJ...833....3A, 2017arXiv171001191I,2018Sci...361..147I} and real time alerts{\footnote {\url{https://gcn.gsfc.nasa.gov/amon.html}}; see also {\url{ https://icecube.wisc.edu/science/data/TXS0506_alerts}}}. The high-energy neutrino candidates come from several different analysis chains: the muon tracks (MUONT), high-energy starting events (HESEA), extremely high-energy alerts (EHEA), and the GOLD and BRONZE events that replaced the HESEA and EHEA chains in mid-2019.
These lists were used as a starting point by \citet{2020ApJ...894..101P} and the above-described considerations led them to perform a cut in positional accuracy and energy. The cuts were performed at the 90\% containment area on the celestial sphere $\Omega_{90}< 10\deg^2$ and E$>200$ TeV, and only track-like events were considered. For EHEA neutrino alerts, energies were not published, but by definition these events have extremely high energy
(the pipeline is optimized for events with E$\geq500$\,TeV)
\citep{2017APh....92...30A} and so we include all of them in our event selection. We follow almost the same criteria as \citet{2020ApJ...894..101P} and therefore the list of neutrinos used in our work is almost identical to theirs.

We have only one additional neutrino event in our list.  This is the event on May 15, 2012 (MJD 56062), exactly at E=200TeV, which was excluded from \citet{2020ApJ...894..101P} because they used a noninclusive cutoff E$>200$ TeV, while we use E$\geq200$\,TeV. 
We did not consider neutrino events after May 2020 because, for these events, analyzing the radio activity around the neutrino event with a reasonable time window would be impossible. We also exclude one event at DEC $<-86^\circ$ because sources at this low declination are not included in OVRO or Mets\"ahovi monitoring. In total, our sample includes 56 neutrino events, which are listed in Table~\ref{table:neut}.

\begin{table*}
\centering
  \caption{IceCube neutrino events at E$\geq200$ TeV used in our analyses} 
\label{table:neut}
  \begin{tabular}{lrlrrrrrrrl}
    \hline
    \hline
  \multicolumn{1}{c}{Date} &
  \multicolumn{1}{c}{MJD} &
  \multicolumn{1}{c}{Type} &
  \multicolumn{1}{c}{RA} &
  \multicolumn{2}{c}{RA Error} &
  \multicolumn{1}{c}{DEC} &
  \multicolumn{2}{c}{DEC Error} &
                                   \multicolumn{1}{c}{Energy} &
                                                                 Reference\\
      \multicolumn{1}{c}{} &
  \multicolumn{1}{c}{} &
  \multicolumn{1}{c}{} &
  \multicolumn{1}{c}{(deg.)} &
  \multicolumn{2}{c}{(deg.)} &
  \multicolumn{1}{c}{(deg.)} &
  \multicolumn{2}{c}{(deg.)} &
                                   \multicolumn{1}{c}{(TeV)} & \\
(1) & (2) & (3) & (4) & (5) & (6) & (7) & (8) & (9) & (10) & (11) \\
    
\hline
  2009-08-13 & 55056 & MUONT & 29.51 & +0.40 & $-0.38$ & 1.23 & +0.18 & $-$0.22 & 480 & 1\\
  2009-11-06 & 55141 & MUONT & 298.21 & +0.53 & $-0.57$ & 11.74 & +0.32 & $-$0.38 & 250 & 1\\
  2010-06-23 & 55370 & MUONT & 141.25 & +0.46 & $-0.45$ & 47.80 & +0.56 & $-$0.48 & 260 & 1\\
  2010-09-25 & 55464 & MUONT & 266.29 & +0.58 & $-0.62$ & 13.40 & +0.52 & $-$0.45 & 460 & 1\\
  2010-10-09 & 55478 & EHEA & 331.09 & +0.56 & $-0.72$ & 11.10 & +0.48 & $-$0.58 & -& 1,2 \\
  2010-10-28 & 55497 & EHEA & 88.68 & +0.54 & $-0.55$ & 0.46 & +0.33 & $-$0.27 & - & 1,2 \\
  2010-11-13 & 55513 & MUONT & 285.95 & +1.29 & $-1.50$ & 3.15 & +0.70 & $-$0.63 & 520 & 1\\
  2011-01-28 & 55589 & EHEA & 307.53 & +0.82 & $-0.81$ & 1.19 & +0.35 & $-$0.32 & - & 1,2\\
  2011-03-04 & 55624 & EHEA & 116.37 & +0.73 & $-0.73$ & $-$10.72 & +0.57 & $-$0.65 & - &2\\
  2011-05-21 & 55702 & MUONT & 235.13 & +2.70 & $-1.76$ & 20.30 & +0.44 & $-$0.62 & 300 & 1\\
  2011-06-10 & 55722 & MUONT & 272.22 & +1.23 & $-1.19$ & 35.55 & +0.28 & $-$0.29 & 210 & 1\\
  2011-07-14 & 55756 & HESEA & 67.86 & +0.51 & $-0.72$ & 40.32 & +0.73 & $-$0.25 & 253 &
  2,3\\
  2011-09-30 & 55834 & EHEA & 266.48 & +2.09 & $-1.55$ & $-$4.41 & +0.59 & $-$0.86 & - & 2\\
  2012-03-01 & 55987 & EHEA & 238.01 & +0.60 & $-0.59$ & 18.60 & +0.46 & $-$0.39 & - & 2\\
  2012-05-15 & 56062 & MUONT & 198.74 & +1.44 & $-1.09$ & 31.96 & +0.81 & $-$0.58 & 200 & 1\\
  2012-05-23 & 56070 & EHEA & 171.03 & +0.81 & $-0.90$ & 26.36 & +0.49 & $-$0.30 & - &2\\
  2012-08-07 & 56146 & MUONT & 330.10 & +0.65 & $-0.82$ & 1.57 & +0.46 & $-$0.42 & 260 & 1\\
  2012-09-22 & 56192 & EHEA & 70.75 & +1.56 & $-1.63$ & 19.79 & +1.37 & $-$0.68 & - &2\\
  2012-10-11 & 56211 & EHEA & 205.22 & +0.59 & $-0.65$ & $-$2.39 & +0.51 & $-$0.57 & - & 1,2\\
  2012-10-26 & 56226 & MUONT & 169.61 & +1.16 & $-1.11$ & 28.04 & +0.67 & $-$0.66 & 750 & 1\\
  2013-06-27 & 56470 & HESEA & 93.43 & +0.80 & $-0.85$ & 14.02 & +0.72 & $-$0.75 & 200 &2\\
  2013-08-17 & 56521 & MUONT & 224.89 & +0.87 & $-1.19$ & $-$4.44 & +1.21 & $-$0.94 & 400 & 1\\
  2013-09-07 & 56542 & EHEA & 129.81 & +0.48 & $-0.28$ & $-$10.36 & +0.36 & $-$0.31 & - &2\\
  2013-10-14 & 56579 & MUONT & 32.94 & +0.63 & $-0.62$ & 10.20 & +0.34 & $-$0.49 & 390 & 1\\
  2013-10-23 & 56588 & EHEA & 301.82 & +1.10 & $-0.93$ & 11.49 & +1.19 & $-$1.09 & - &2\\
  2013-12-04 & 56630 & EHEA & 289.16 & +1.08 & $-0.94$ & $-$14.25 & +0.91 & $-$0.81 & - &2\\
  2014-01-08 & 56665 & EHEA & 344.53 & +0.67 & $-0.48$ & 1.57 & +0.35 & $-$0.32 & - &2\\
  2014-01-09 & 56666 & EHEA & 292.85 & +0.87 & $-0.94$ & 33.06 & +0.50 & $-$0.46 & - & 1,2\\
  2014-02-03 & 56691 & EHEA & 349.54 & +2.21 & $-1.97$ & $-$13.71 & +1.23 & $-$1.38 & - &2\\
  2014-06-09 & 56817 & MUONT & 106.26 & +2.27 & $-1.90$ & 1.29 & +0.83 & $-$0.74 & 340 & 1\\
  2014-06-11 & 56819 & EHEA & 110.30 & +0.66 & $-0.45$ & 11.57 & +0.14 & $-$0.24 & - & 1,2\\
  2014-09-23 & 56923 & EHEA & 169.72 & +0.91 & $-0.86$ & $-$1.34 & +0.73 & $-$0.66 & - &2\\
  2015-01-27 & 57049 & MUONT & 100.48 & +0.95 & $-1.87$ & 4.56 & +0.68 & $-$0.50 & 210 & 1\\
  2015-05-15 & 57157 & MUONT & 91.60 & +0.16 & $-0.74$ & 12.18 & +0.37 & $-$0.35 & 240 & 1\\
  2015-07-14 & 57217 & MUONT & 325.50 & +1.77 & $-1.46$ & 26.10 & +1.68 & $-$1.85 & 300 & 4\\
  2015-08-12 & 57246 & EHEA & 328.19 & +1.01 & $-1.03$ & 6.21 & +0.44 & $-$0.49 & - & 2,5\\
  2015-08-31 & 57265 & EHEA & 54.85 & +0.94 & $-0.98$ & 33.96 & +1.07 & $-$1.19 & - &2\\
  2015-09-04 & 57269 & MUONT & 134.00 & +0.39 & $-0.58$ & 28.00 & +0.47 & $-$0.47 & 220 &5\\
  2015-09-23 & 57288 & EHEA & 103.27 & +0.70 & $-1.36$ & 3.88 & +0.59 & $-$0.71 & - &2\\
  2015-09-26 & 57291 & EHEA & 194.50 & +0.76 & $-1.21$ & $-$4.34 & +0.70 & $-$0.95 & - &2\\
  2015-11-14 & 57340 & MUONT & 76.30 & +0.75 & $-0.74$ & 12.60 & +0.61 & $-$0.58 & 740 &5\\
  2015-11-22 & 57348 & EHEA & 262.18 & +0.90 & $-1.21$ & $-$2.38 & +0.73 & $-$0.43 & - &2\\
  2016-01-28 & 57415 & EHEA & 263.40 & +1.35 & $-1.18$ & $-$14.79 & +0.99 & $-$1.02 & - &2\\
  2016-03-31 & 57478 & MUONT & 15.60 & +0.45 & $-0.58$ & 15.60 & +0.53 & $-$0.60 & 380 & 4\\
  2016-05-10 & 57518 & EHEA & 352.34 & +1.63 & $-1.31$ & 2.09 & +0.99 & $-$0.85 & - &2\\
  2016-07-31 & 57600 & EHEA & 214.50 & +0.75 & $-0.75$ & $-$0.33 & +0.75 & $-$0.75 & - &2,5\\
  2016-08-06 & 57606 & EHEA & 122.81 & +0.50 & $-0.50$ & $-$0.81 & +0.50 & $-$0.50 & - &2\\
  2016-12-10 & 57732 & EHEA & 46.58 & +1.10 & $-1.00$ & 14.98 & +0.45 & $-$0.40 & - &2\\
  2017-03-21 & 57833 & EHEA & 98.30 & +1.20 & $-1.20$ & $-$15.02 & +1.20 & $-$1.20 & - &2\\
  2017-09-22 & 58018 & EHEA & 77.43 & +0.95 & $-0.65$ & 5.72 & +0.50 & $-$0.30 & 290 &6\\
  2017-11-06 & 58063 & EHEA & 340.00 & +0.70 & $-0.50$ & 7.40 & +0.35 & $-$0.25 & - &7\\
  2018-09-08 & 58369 & EHEA & 144.58 & +1.55 & $-1.45$ & $-$2.13 & +0.90 & $-$1.20 & - &7\\
  2018-10-23 & 58414 & EHEA & 270.18 & +2.00 & $-1.70$ & $-$8.57 & +1.25 & $-$1.30 & - &7\\
  2019-03-31 & 58573 & HESEA & 337.68 & +0.23 & $-$0.34 & $-$20.70 & +0.30 & $-$0.48 & 1987 &7\\
  2019-05-03 & 58606 & EHEA & 120.28 & +0.57 & $-0.77$ & 6.35 & +0.76 & $-$0.70 & - &7\\
  2019-07-30 & 58694 & GOLD & 225.79 & +1.28 & $-1.43$ & 10.47 & +1.14 & $-$0.89 & 299 &7\\
\hline\end{tabular}
\tablebib{(1)~\cite{2016ApJ...833....3A}; (2)~\cite{alertevents}; (3)~\cite{2015arXiv151005223T}; (4)~\cite{2019arXiv190711699I}; (5)~\cite{2017arXiv171001191I}; (6)~\cite{2018Sci...361..147I}; (7)~\tiny\url{https://gcn.gsfc.nasa.gov/amon.html} }
\end{table*}

\subsection{OVRO data and samples} \label{ovro}
The OVRO blazar monitoring program \citep{2011ApJS..194...29R} has been  continuously ongoing since 2008. It uses the 40~m Telescope at the Owens Valley Radio Observatory (37 d 14$\arcmin$ 02$\arcsec$ N, 118 d 16$\arcmin$ 56$\arcsec$ W). The OVRO 40~m uses two symmetric off-axis beams and a cryogenic receiver with a 15.0~GHz center frequency and 3~GHz bandwidth. The source is alternated between the two beams in an ON-ON fashion to remove atmospheric and ground contamination. The receiver was replaced in May 2014 with a dual-beam correlation receiver instead of the original Dicke switched receiver, providing better performance and higher time stability. Calibration is achieved using a temperature-stable diode noise source to remove receiver gain
drifts and the flux density scale is determined from observations of
3C~286 assuming the \cite{baars77} value of 3.44~Jy at
15.0~GHz. The systematic uncertainty of about 5\% in the flux density scale is not included in the error bars.  Complete details of the reduction and calibration procedure are given in \cite{2011ApJS..194...29R}.

Originally the monitoring included only sources from the CGRaBS sample, which is a catalog of AGNs with radio and X-ray properties similar to EGRET $\gamma$-ray blazars \citep{healey08}. CGRaBS sample objects are derived from a parent population of flat-spectrum radio sources ($\alpha > -0.5$) with 4.8\,GHz flux density $> 65$\,mJy. In the OVRO monitoring, 1158 sources at Dec $> -20^\circ$ were included. One faint source CGRaBS~J1310+3233 was later removed from the sample because it is exactly 13 arcmin distant from another bright CGRaBS target, CGRaBS~J1310+3220, which corresponds to the separation of the two beams of the receiver, resulting in confusion in its fluxes. The final CGRaBS sample used in this paper consists of 1157 objects (hereafter CGRaBS sample).

In addition to the CGRaBS sample, the OVRO monitoring sample has subsequently been increased by adding $\gamma$-ray-detected sources from the first $Fermi$-LAT AGN catalog (1LAC, \citealt{abdo10b}) and 2LAC \citep{ackermann11}. Additionally, the source list includes various other AGNs and blazars "of interest", meaning that the total number of monitored AGNs is currently 1795 (hereafter the `all-AGN' sample).  We note that the full OVRO sample is not statistically well defined, and so our analysis that uses it should be treated with caution, and we do not use it when assessing the significance of our results. We return to this point in our Discussion in Sect.~\ref{discussion}.

Based on a source-count distribution, \cite{liodakis17} determined that the OVRO all-AGN sample is complete down to a flux density limit of 350\,mJy when using the maximum likelihood mean flux density at 15\,GHz from \cite{richards14}. 
The sample in \cite{richards14} includes only objects in the CGRaBS catalog and the 1LAC catalog at DEC$>-20^\circ$, and excludes objects near the Galactic plane with Galactic latitude $|b|<10^\circ$. There are 589 objects with maximum likelihood mean flux density greater than 350\,mJy forming the flux-density-limited sample (hereafter OVRO-350mJy sample). However, we note that because the source-count distribution is not derived from a full-sky survey, there are still individual sources brighter than this limit missing from our sample.

\cite{2020ApJ...894..101P} used the RFC catalog down to its completeness limit of 150\,mJy of unresolved flux density at 8\,GHz in their analysis (completeness also determined from a source-count distribution). For direct comparison with their results, we cross-matched our full sample with the RFC and use the resulting 1156 objects in our sample (hereafter RFC-150mJy sample). Because the first IceCube neutrino event meeting our selection criteria is from 2009, in our analysis we use data from January 1, 2008, until July 1, 2020.

\subsection{Mets\"ahovi data and sample}
A sample of over 1000 AGNs is being monitored at Aalto University Mets\"ahovi Radio Observatory in Finland (60 d 13$\arcmin$ 04$\arcsec$ N, 24 d 23$\arcmin$ 35$\arcsec$ E).
The monitored sample is heterogeneous. The current high-priority list contains over 400 sources where the aim is to observe them regularly, and weekly observations are being made for about 100 sources, for some of which the data sets run for over 40 years. At the flux density limit of 2\,Jy at DEC$>0^\circ$, the sample can be considered complete \citep{valtaoja92}.

The Mets\"ahovi 14-m telescope has a radome and Cassegrain optics.
The measurements were made with a 1 GHz-band dual beam
receiver centered at 36.8 GHz. The HEMT (high electron mobility
transistor) front end operates at ambient temperature. The
observations are Dicke switched ON--ON observations, alternating the
source and the sky in each feed horn. A typical integration time to
obtain one flux density data point is between 1200 and 1600 s. The
detection limit of the telescope at 37 GHz is on the order of 0.2 Jy
under optimal conditions, but is highly weather-dependent. 
Data points with a signal-to-noise ratio of $<4$ are handled as nondetections.

The flux density scale is set by observations of the HII region DR21. 
Sources NGC 7027, 3C 274, and 3C 84 are used as secondary
calibrators. A detailed description of the data reduction and analysis
is given in \cite{terasranta98}. The error estimate in
the flux density includes the contribution from the measurement rms
and the uncertainty of the absolute calibration.

We selected sources that have at least 6 years of data between January 1, 2008, and February 29, 2020. Additionally, we require that there be at least ten detections (signal-to-noise ratio $\geq4$) during that time period. We also applied a cut on the flux density to exclude sources that are too close to the detection threshold, and only included sources that had a maximum flux $>0.7$\,Jy. Applying these criteria to the  Mets\"ahovi database resulted in 183 sources. The source with the lowest declination is at DEC$=-13.08^\circ$, which means that there are two neutrino events that are excluded from the comparison with the Mets\"ahovi sample, when the statistical uncertainties and the maximum systematic uncertainty of $1^\circ$ in the neutrino event positions  (see Sect.~\ref{positions}) is accounted for.

\citet{2020ApJ...894..101P} found the strongest temporal relation between neutrino arrival times and radio fluxes in their highest frequency observations at 22\,GHz. Because of opacity effects, the 37\,GHz (and 22\,GHz) emission originates closer to the central black hole than the 15\,GHz emission. Although the Mets\"ahovi sample is not statistically complete, it is interesting to compare the results of these well-sampled light curves at 37\,GHz with the OVRO 15\,GHz. However, we note that the results from the Mets\"ahovi sample alone should only be treated as indicative because of the incompleteness of the parent sample.

\section{Analyses and results}\label{analyses}

We looked at three different metrics when studying the association of IceCube neutrinos with radio data. 
\begin{enumerate}
    \item Mean flux density of the associated sources. 
    \item Mean activity index of the associated sources.
    \item Number of flaring sources in the sample.
\end{enumerate}
In Sect.~\ref{positions}, we describe how we match the neutrino positions with our radio samples. Following \cite{2020ApJ...894..101P}, we estimate the (unknown) systematic uncertainties in the IceCube event positions using our data. We also describe how we generate random Monte Carlo samples to obtain the chance coincidence probability in our analysis. In Sect.~\ref{meanflux} we discuss the analysis of the mean flux density, while the mean activity index and the number of flaring sources in the sample are discussed in Sect.~\ref{timedependent}.

\subsection{Position matching and systematic uncertainty estimation}\label{positions}
\begin{figure}
   \centering
   \includegraphics[width=\hsize]{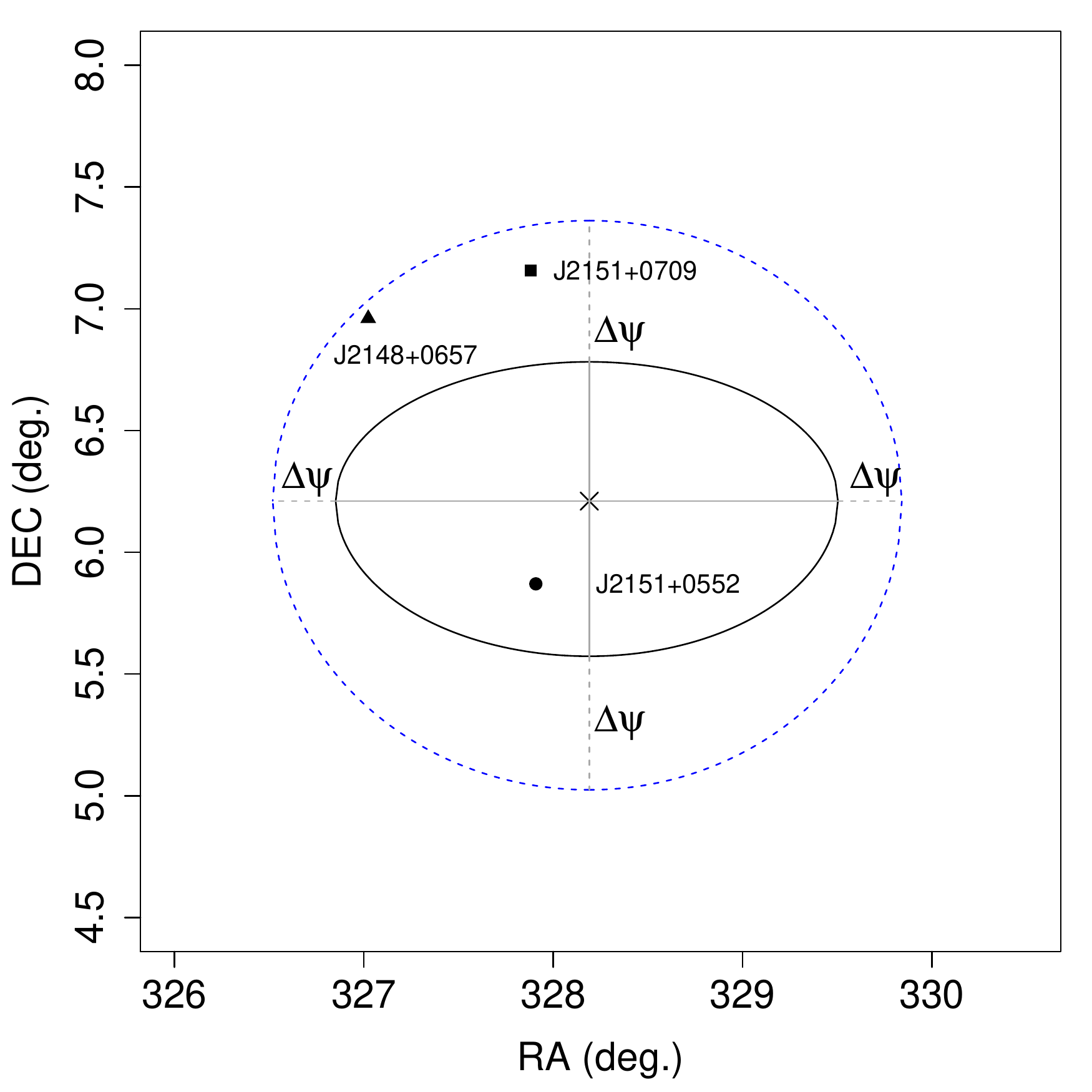}
      \caption{ Example of the elliptical uncertainty region around the neutrino event on August 12, 2015 (MJD 57246; shown with a black cross), and the positions of the three associations found for it in the OVRO sample. The gray solid lines mark the asymmetric statistical uncertainties in each direction (given in Table~\ref{table:neut}), multiplied by the factor of 1.3 discussed in the text. The solid black line shows the two-dimensional error ellipse when only statistical uncertainties are included. The gray dashed lines show the additional systematic uncertainty $\Delta \psi = 1.0 ^\circ$ as found optimal for the mean flux density analysis in the CGRaBS sample, which is added in quadrature to the statistical uncertainties in all directions, and the blue dashed line marks the resulting two-dimensional error ellipse.}
         \label{ellipse}
   \end{figure}

 The point-spread function of IceCube can be approximated as a two-dimensional Gaussian \citep{IceCube2014}, but the statistical uncertainties of the published events (Table~\ref{table:neut} columns 5, 6, 8, and 9) are given as 90\% containment intervals in right ascension (RA) and declination (DEC) separately, and they can also be asymmetric. When translated  to an ellipse on the sky, the two-dimensional containment region is therefore less than 90\%. This means that we must first translate the coordinate-wise errors into two-dimensional coverage regions. This is done by multiplying the individual statistical errors in RA and DEC with the ratio of 90\% quantiles of two- and one-dimensional Gaussian distributions: $\frac{\sqrt{-\log(1-0.9)}}{\mathrm{erf}^{-1}(0.9)}\approx 1.3$ \citep{2020ApJ...894..101P}. This way we obtain regions around the neutrino events that are bounded by four quarters of ellipses (see Fig.~\ref{ellipse}).

In addition to the statistical errors, there is also an additional systematic positional uncertainty, which is dominated by the uncertainties in the optical properties of the ice \citep{2013Sci...342E...1I}. The systematic uncertainties are typically not published for the individual events as their estimation requires a more detailed Monte Carlo reconstruction of the events, but the upper limit for track-like events is $1.0^\circ$ based on studies of the cosmic-ray shadow of the Moon by the IceCube Collaboration \citep{IceCube2014}. Recently, the IceCube Collaboration discussed an updated analysis of the high-energy starting events \citep{2020arXiv201103545A}, noting that an improved uncertainty estimation resulted on average in uncertainties that are nearly twice as large as previously reported. It is therefore important that this unknown uncertainty is properly accounted for. 

Following \cite{2020ApJ...894..101P}, we estimate the unknown additional systematic uncertainty in the neutrino positions using our data.  As discussed in \cite{2020ApJ...894..101P}, this is motivated by a commonly used approach in high-energy astrophysics, where multiple values of an unknown parameter are tried to select the one with the strongest signal. We leave the systematic uncertainty parameter $\Delta \psi$ free in our analysis, and use it as a trial value. We scan over a range from $0.1^\circ$ to $1.0^\circ$ with a step of $\Delta \psi = 0.1^\circ$, and add $\Delta \psi$ in quadrature to the statistical uncertainties in RA and DEC in all directions so that our final uncertainty in each direction is $\sqrt{\Delta_\mathrm{stat}^2 + \Delta\psi^2}$.

This is slightly different to \cite{2020ApJ...894..101P} who added the systematic uncertainties directly to the statistical ones, and our final uncertainty regions are therefore smaller than the ones they used, which has an effect on the associations we find. We note that our main conclusions are not affected by this difference.

For each $\Delta \psi$ we first find all the sources in our samples that fall within the elliptical region around the neutrino event (see Fig.~\ref{ellipse} for an example). We then calculate the test statistic of interest (mean flux density or activity index) for each source. We repeat this step for random Monte Carlo samples. Similar to \cite{2020ApJ...894..101P}, we generate random comparison samples by shifting the IceCube neutrino positions in RA while keeping the DEC constant. This way we reproduce the effect that the sensitivity of the instrument depends only on the zenith angle \citep{2017ApJ...835..151A}. We then match these random neutrino samples with our observed samples, and calculate the same test statistic (mean flux density and activity index) for each associated random source. 

\begin{figure}
   \centering
   \includegraphics[width=\hsize]{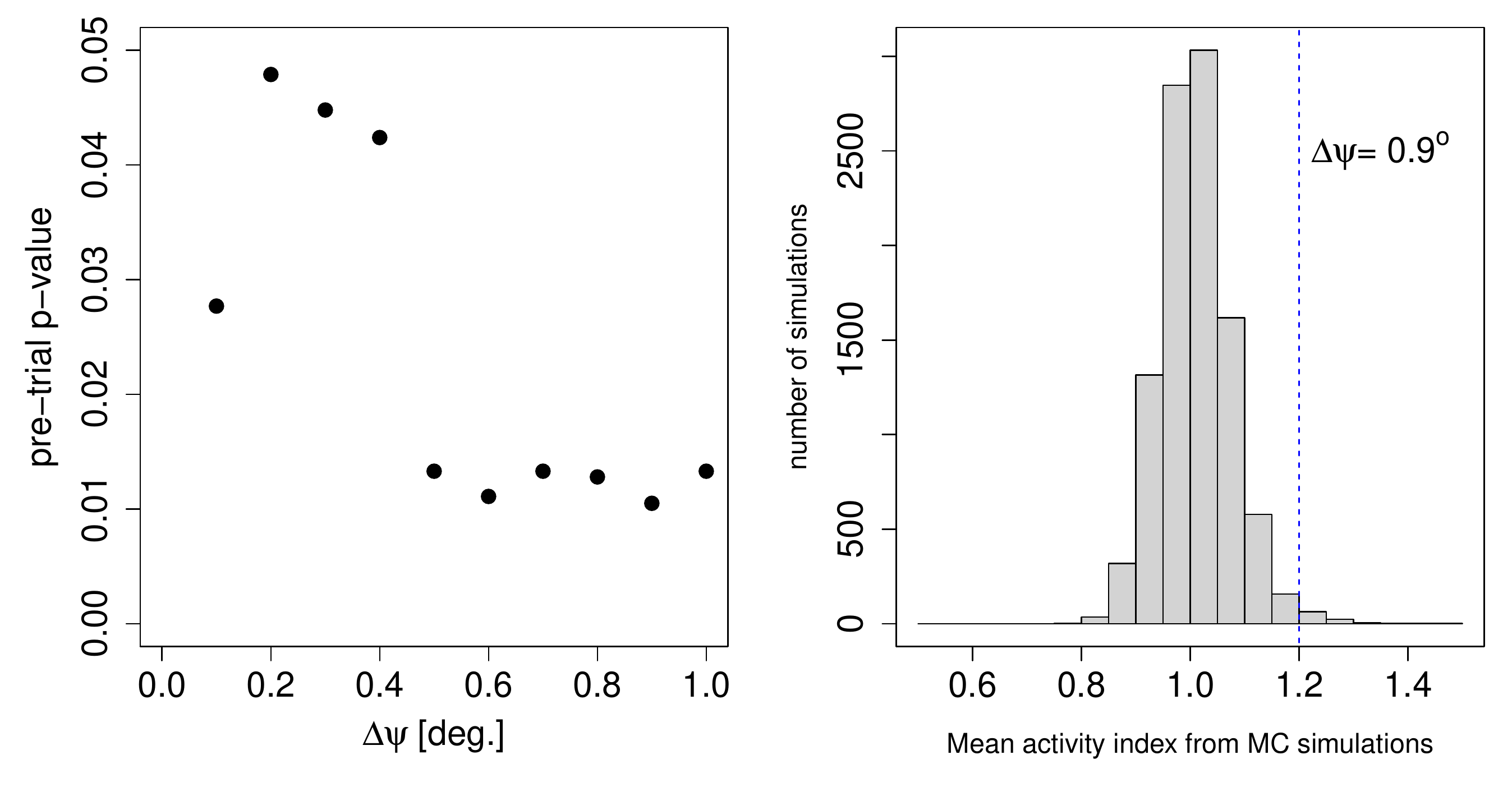}
      \caption{ Example of the Monte Carlo process for estimating the pre-trial p-values. Left: Pre-trial p-value for different $\Delta \psi$ values in the CGRaBS sample mean activity index analysis (Sect.~\ref{meanai}). The minimum pre-trial p-value of 0.010 is obtained for $\Delta \psi=0.9^\circ$. Right: Distribution of the mean activity indices from the random samples for $\Delta \psi=0.9^\circ$. The dashed blue line indicates the mean activity index of the associations in the real CGRaBS sample.}
         \label{MCexample}
   \end{figure}

With this procedure we can estimate the distribution of the test statistic under the null hypothesis of no association between the neutrino events and the radio sources. We can then estimate, under the null hypothesis, the chance probability of obtaining a value of the test statistic equal to or larger than the one obtained from the data as,
\begin{equation}\label{eq:pval}
    p = \frac{M+1}{N+1},
\end{equation}
where $M$ is the number of random samples with a larger test statistic than in the real data, and $N$ is the number of random samples \citep{davison13}, which in our case is 10~000. 
This is given as the pre-trial p-value when reporting our results. We then select as the optimal systematic uncertainty parameter the value of $\Delta \psi$ that gives the smallest pre-trial p-value.  An example of this process is shown in Fig.~\ref{MCexample}.

In order to account for the factor resulting from these multiple trials of $\Delta \psi$, we again follow \cite{2020ApJ...894..101P} and insert the calculation of the post-trial p-value into our Monte Carlo simulations. To do this, we treat each random sample as the real observation,  and compare it against the other 9999 random samples. As a result, we first obtain a p-value for each trial value of $\Delta \psi$ in a completely random population, and as for the real data  can similarly select the smallest pre-trial p-value. We then calculate how many times we obtain a pre-trial p-value in these random populations that is as small as that obtained for the real data. This value is used as $M$ in Eq.~\ref{eq:pval}, while N will be 9999. This gives us the post-trial p-value, which accounts for the multiple trials of $\Delta \psi$.

Because we are using multiple subsets of the OVRO sample and the Mets\"ahovi sample to calculate the test statistics, we also need to account for multiple trials arising from these tests. The most conservative approach is to multiply the resulting p-values by the number of samples (five in our case) to obtain the final post-trial p-value. We note that, because our samples are not independent, the true trial factor is smaller than that \citep[e.g.,][]{tinyakov04}.  We tested this by using Monte Carlo simulations to estimate the trial factor when considering all the samples together, and the main conclusions of the paper remain the same. Thus, in the following we use the most conservative approach and multiply the p-values by the factor of five.

\subsection{Mean flux density}\label{meanflux}
The mean flux densities of the associated sources in the different samples  calculated over the entire duration of the light curve are given in Table~\ref{table:ovro} for the OVRO samples and Table~\ref{table:metsahovi} for the Mets\"ahovi sample.  The optimal systematic uncertainty parameter $\Delta \psi$ was $1.0^\circ$ for the OVRO samples and $0.9^\circ$ for the Mets\"ahovi sample. These are slightly larger than the values reported in \cite{2020ApJ...894..101P}, but the difference can be explained by our different way of adding the systematic uncertainty to the statistical ones. The optimal value and the number of associations within this limit are given in Table~\ref{table:meanflux}.  When comparing the mean flux density values with random samples, we average the values of the individual sources within the sample of associated sources to obtain a single parameter for the sample. These are also given in Table~\ref{table:meanflux}, along with the resulting pre-trial and post-trial p-values. The OVRO and Mets\"ahovi light curves of these sources are shown in Appendix A.

\begin{table*}
\caption{Mean flux density (Sect. \ref{meanflux}) of OVRO AGNs within the error regions of IceCube events}
\label{table:ovro}
\begin{tabular}{ll|llrrllllr} 
\hline
\hline
\multicolumn{2}{c}{IceCube Event} & \multicolumn{9}{c}{AGN} \\
MJD & Category & J2000 Name & Other Name & RA & DEC &  all & CGR & 350 & 150 & $<S>$ \\ 
 &  &  &  & (deg.) & (deg.) &  & & & & (Jy)  \\ 
(1) & (2) & (3) & (4) & (5) & (6) & (7) & (8) & (9) & (10) & (11) \\ 
\hline
55478.38058 & EHEA & J2203+1007 & B2201+098 & 330.8790 & 10.1283 &  X & X & $\dotsc$  & X & 0.13\\ 
& & J2200+1030 & TXS~2157+102 & 330.0330 & 10.5021 & X & $\dotsc$ &$\dotsc$  & X & 0.22\\ 
\hline
55834.44514 & EHEA & J1743$-$0350 & PKS~1741$-$038 & 265.9950 & $-$3.8344 & X &X  & X  & X & 4.24\\ 
\hline
 56062.96000 & MUONT & J1310+3220 & B1308+326 & 197.6190 & 32.3453 & X & X & X  & X & 2.08\\ 
\hline
  56070.57428 & EHEA & J1125+2610 & PKS~1123+264 & 171.4740 & 26.1722 & X & X & X  & X & 0.88\\ 
\hline
 56146.21000 & MUONT & J2200+0234 & GB6~B2158+0219 & 330.2060 & 2.5750 & X & X & $\dotsc$ & $\dotsc$ & 0.10\\ 
\hline
 56579.91000 & MUONT & J0211+1051 & GB6~B0208+1037 & 32.8049 & 10.8594 & X & X & X  & X & 0.90\\ 
\hline
  56630.47006 & EHEA & J1916$-$1519 & B1914$-$154 & 289.2190 & $-$15.3167 & X & X  & $\dotsc$  & X & 0.21\\ 
\hline
  57246.75909 & EHEA & J2151+0552 & PKS~2149+056 & 327.9080 & 5.8703 & X &X  &X   &X  & 0.40\\ 
 & & J2148+0657 & PKS~2145+067 & 327.0230 & 6.9606 & X & X & X & X & 3.99\\ 
 &  & J2151+0709 & PKS~2149+069 & 327.8810 & 7.1572 & X & X & X & X & 0.84\\ 
\hline
  57265.21782 & EHEA & J0341+3352 & TXS~0337+337 & 55.2916 & 33.8725 & X &X  & $\dotsc$  & $\dotsc$ & 0.10\\ 
& & J0336+3218\tablefootmark{a} & B0333+321 & 54.1254 & 32.3082  & X &$\dotsc$  & $\dotsc$  & X & 1.86 \\ 
\hline
  57291.90119 & EHEA & J1256$-$0547 & 3C279 & 194.0470 & $-$5.7892 & X & X & X  & X & 18.02\\ 
\hline
  57340.90000 & MUONT & J0502+1338 & PKS~0459+135 & 75.6384 & 13.6361 & X  &   X & X & X  & 0.44\\ 
\hline
  57833.31413 & EHEA & J0631$-$1410 & TXS~0629$-$141 & 97.8343 & $-$14.1753 &  X & X & X  & X & 0.59\\ 
& & J0630$-$1323 & TXS~0628$-$133 & 97.7246 & $-$13.3928  & X & X & X  & X & 0.43\\ 
\hline
  58018.87118 & EHEA & J0509+0541 & TXS~0506+056 & 77.3582 & 5.6931 & X &X  &X   & X & 0.69\\ 
\hline
  58063.65254 & EHEA & J2238+0724 & GB6~B2235+0708 & 339.5430 & 7.4039 & X &X  & $\dotsc$  &X  & 0.16\\ 
\hline
  58694.74353 & GOLD & J1504+1029 & PKS~1502+106 & 226.1040 & 10.4942 & X & X & X  & X & 2.07\\ 
\hline
\end{tabular}
\tablefoot{Cols (5) and (6) give the coordinates of the associated AGNs. Cols 7-10 indicate the sample where the match is found. `All'  stands for all-AGN, `CGR' for CGRaBS, `350' for OVRO-350mJy, and `150' for
  RFC-150mJy samples. The optimal systematic uncertainty was
  $1.0^\circ$ for all the samples. Col (11) gives the mean flux density of the source.}
  \tablefoottext{a}{Although this source is brighter than 350 mJy, it is not included in the OVRO-350mJy sample, because it is not included in the sample of \cite{richards14}, which we use to derive our flux-density limited sample.}
\end{table*}

\begin{table*}
\caption{Mean flux density (Sect. \ref{meanflux}) and activity index (Sect. \ref{timedependent}) of Mets\"ahovi AGN within the error regions of IceCube events} 
\label{table:metsahovi}
\begin{tabular}{ll|llrrrr} 
\hline
\hline
\multicolumn{2}{c}{IceCube Event} & \multicolumn{6}{c}{AGN} \\
MJD & Category & J2000 Name & Other Name & RA & DEC & $<S>$ & A.I. \\ 
 &  &  &  & (deg.) & (deg.) & (Jy) &   \\ 
(1) & (2) & (3) & (4) & (5) & (6) & (7) & (8) \\ 
\hline
55834.44514 & EHEA & J1743$-$0350 & PKS~1741$-$038 & 265.9953 & $-$3.8346 & 3.61 & \textbf{1.36}\\ 
\hline
56062.96000 & MUONT & J1310+3220 & B1308+326 & 197.6194 & 32.3455 & 1.76 & \textbf{1.28}\\ 
 &  & J1310+3233\tablefootmark{a} & B1308+328 & 197.7475 & 32.5596 & 0.49 & 0.81\\ 
\hline
56579.91000 & MUONT & J0211+1051 & GB6~B0208+1037 & 32.8049 & 10.8597 & 1.15 & \textbf{2.04}\\ 
\hline
57291.90119 & EHEA & J1256$-$0547 & 3C279 & 194.0465 & $-$5.7893 & 19.07 & $\dotsc$\\ 
\hline
58018.87118 & EHEA & J0509+0541 & TXS~0506+056 & 77.3579 & 5.6931 & 1.17 & 0.42\\ 
\hline
58694.74353 & GOLD & J1504+1029 & PKS1502+106 & 226.1041 & 10.4942 & 2.05 & \textbf{1.69}\\ 
\hline
\end{tabular}
\tablefoot{Cols (5) and (6) give the coordinates of the associated AGNs. Col (7) gives the mean flux density and col (8) the activity index of the source. The optimal systematic uncertainty was $0.9^\circ$ for the mean flux density analysis and $0.5^\circ$ for the activity index analysis, which is why no A.I. is given for J1256$-$0547 as it is outside this range. Sources marked with bold-face font in Col. 8 are those showing strong flaring activity (Sect. \ref{flaring})}
\tablefoottext{a}{This source is the only one of the Mets\"ahovi associations not included in the OVRO samples. It is included in the 4LAC catalog and its optical class is FSRQ and SED class LSP.}
\end{table*}

\begin{table*}
\caption{Chance coincidence for the mean flux density analysis}             
\label{table:meanflux}      
\centering          
\begin{tabular}{l l l l l l l}     
\hline\hline      
Sample & N$_\mathrm{Source}$  & $\Delta \psi$ & N$_\mathrm{Assoc}$  & $<S>$ & pre-trial & post-trial \\
 &  & (deg) &  & (Jy) & p & p \\
 (1) & (2) & (3) & (4) & (5) & (6) & (7) \\
\hline                    
OVRO all-AGN & 1795 & 1.0 & 20 & 1.92 & 0.019 & 0.037$(\times5)$  \\  
CGRaBS & 1157 & 1.0 & 18 & 2.02 & 0.027 & 0.049$(\times5)$\\
OVRO-350mJy & 589 & 1.0 &  13 & 2.74     & 0.087 & 0.160$(\times5)$ \\
RFC-150mJy & 1156 & 1.0 & 18   &  2.12 & 0.059 & 0.106$(\times5)$ \\
Mets\"ahovi & 183 & 0.9 & 7    & 4.19 & 0.080 & 0.138$(\times5)$ \\
\hline                  
\end{tabular}
\tablefoot{Col (1) indicates the sample studied and col (2) gives the number of sources in that sample. Col (3) gives the optimal systematic uncertainty parameter for the sample, and col (4) the number of associations found using this systematic uncertainty. The pre-trial and post-trial p-values are given in cols (6) and (7), respectively.  The additional trial factor due to multiple samples is shown as a multiplicative factor in Col (7) (see Sect.~\ref{positions}).}
\end{table*}

\subsection{Time-dependent analysis}\label{timedependent}
\subsubsection{Mean activity index}\label{meanai}
Because both OVRO and Mets\"ahovi light curves are well sampled, we can estimate whether any of the associated blazars show flaring behavior during the arrival time of the neutrino. Following \cite{2020ApJ...894..101P}, we use the activity index (A.I.), defined as the mean flux density within a time window around the neutrino event divided by the mean flux density outside this time window. \cite{2020ApJ...894..101P} set the time window as a free parameter similar to the systematic uncertainty; they then selected the time window maximizing the mean activity index of their real sample as the optimal time window, which they found to be 0.9 yr for the RATAN-600 sample.

We explored this approach and find that it does not work for either the OVRO or the Mets\"ahovi data.  The reason is that with the well-sampled light curves from OVRO and Mets\"ahovi, leaving the length of the time window free causes a few individual blazars with large flares around a neutrino event to dominate the optimal time window size. For example, in the OVRO data for TXS~0506+056 (J0509+0541) the optimal time window ($\Delta T$) maximizing the activity index is 5.6 yr, simply because at this limit, the window around the neutrino event (calculated as $\pm\Delta T/2$) includes the peak of the large flare following the neutrino arrival (see Fig.~\ref{Figpage1}). We also found that the pre-trial p-value is very insensitive to the length of the time window with values between 0.8\% and 1.0\% for all window sizes between 0.5 and 3 yr for the CGRaBS sample.

Because the optimal window size $\Delta T$ for each source clearly depends on how the neutrino event is positioned with respect to large flares, we think that $\Delta T$ is more related to the typical  duration of flares of blazars in radio wavelengths than directly to the possible link between neutrinos and radio flares.  The typical duration of radio flares can be estimated, for example, using a structure function analysis as done by \cite{hovatta07}, who also showed that the duration of flares changes with radio frequency. Therefore, we set the time window $\Delta T$ to the median structure function timescale from \cite{hovatta07}, which is 2.3 yr at 15\,GHz and 1.4 yr at 37\,GHz. In our analysis, we calculate the activity index using a time range of $\pm\Delta T/2$ around the neutrino event.

For the activity index analysis we find the optimal systematic uncertainty parameter $\Delta \psi$ to be $0.9^\circ$ for the OVRO samples and $\Delta \psi = 0.5^\circ$ for the Mets\"ahovi sample. The activity index for the associated sources in our samples are tabulated in Tables \ref{table:ovro_ai} and \ref{table:metsahovi} for the OVRO and Mets\"ahovi samples, respectively. 

\begin{table*}
\caption{Activity index (Sect. \ref{timedependent}) of OVRO AGN within the error regions of IceCube events} 
\label{table:ovro_ai}
\begin{tabular}{ll|llrrllllr} 
\hline
\hline
\multicolumn{2}{c}{IceCube Event} & \multicolumn{9}{c}{AGN} \\
MJD & Category & J2000 Name & Other Name & RA & DEC & all & CGR & 350 & 150 & A.I. \\ 
 &  &  &  & (deg.) & (deg.) & & & & &   \\ 
(1) & (2) & (3) & (4) & (5) & (6) & (7) & (8) & (9) & (10) & (11) \\ 
\hline
55478.38058 & EHEA & J2200+1030\tablefootmark{a} & TXS~2157+102 & 330.0330 & 10.5021 & X & $\dotsc$ & $\dotsc$  & X & $\dotsc$ \\ 
 &  & J2203+1007 & B2201+098 & 330.8790 & 10.1283 & X & X & $\dotsc$  & X & 0.98\\ 
\hline
  55834.44514 & EHEA & J1743$-$0350 & PKS~1741$-$038 & 265.9950 & $-$3.8344 & X & X & X  & X & 1.20\\ 
\hline
  56062.96000 & MUONT & J1310+3220 & B1308+326 & 197.6190 & 32.3453 & X & X & X  & X & \textbf{1.46}\\ 
\hline
  56070.57428 & EHEA & J1125+2610 & PKS~1123+264 & 171.4740 & 26.1722 & X & X & X  &X  & 1.02\\ 
\hline
56146.21000 & MUONT & J2200+0234 & GB6~B2158+0219 & 330.2060 & 2.5750 & X & X & $\dotsc$ & $\dotsc$ & 0.85\\
\hline
  56579.91000 & MUONT & J0211+1051 & GB6~B0208+1037 & 32.8049 & 10.8594 & X & X & X  &X  & \textbf{1.85}\\ 
\hline
  56630.47006 & EHEA & J1916$-$1519 & B1914$-$154 & 289.2190 & $-$15.3167 & X & X & $\dotsc$  &X  & 0.84\\ 
\hline
  57246.75909 & EHEA & J2151+0709 & PKS~2149+069 & 327.8810 & 7.1572 & X & X & X  &X  & 0.84\\ 
 & & J2151+0552 & PKS~2149+056 & 327.9080 & 5.8703 & X & X & X  & X & 1.03\\ 
\hline
  57265.21782 & EHEA & J0341+3352 & TXS~0337+337 & 55.2916 & 33.8725 & X & X & $\dotsc$  & $\dotsc$ & \textbf{1.29}\\ 
\hline
  57291.90119 & EHEA & J1256$-$0547 & 3C279 & 194.0470 & $-$5.7892 & X & X & X  &X  & 1.17\\ 
\hline
  57340.90000 & MUONT & J0502+1338 & PKS~0459+135 & 75.6384 & 13.6361 & X &X  & X  & X & \textbf{1.63}\\ 
\hline
  57833.31413 & EHEA & J0631$-$1410 & TXS~0629$-$141 & 97.8343 & $-$14.1753 & X &X  &X   &X  & 0.95\\ 
 & & J0630$-$1323 & TXS~0628$-$133 & 97.7246 & $-$13.3928 & X & X & X  & X & \textbf{1.49}\\ 
\hline
  58018.87118 & EHEA & J0509+0541 & TXS~0506+056 & 77.3582 & 5.6931 & X & X & X  & X & 0.99\\ 
\hline
  58063.65254 & EHEA & J2238+0724 & GB6~B2235+0708 & 339.5430 & 7.4039 & X & X & $\dotsc$  & X & 0.96\\ 
\hline
  58694.74353 & GOLD & J1504+1029 & PKS~1502+106 & 226.1040 & 10.4942 & X & X & X  & X & \textbf{1.81}\\ 
\hline
\end{tabular}
\tablefoot{Cols (5) and (6) give the coordinates of the associated AGNs. Cols 7-10 indicate the sample where the match is found. `All' stands for all-AGN, `CGR' for CGRaBS, `350' for OVRO-350mJy, and `150' for RFC-150mJy samples. The optimal systematic uncertainty $\Delta\psi$ was $0.9^\circ$ for all the samples. Sources for which the activity index is marked with bold-face font in Col. 11 are those showing strong flaring activity (Sect. \ref{flaring}).}
\tablefoottext{a}{This source was added to OVRO monitoring only in December 2011 after its inclusion in the 2LAC catalog, which is why there is no activity index available for it around the neutrino event in October 2010.}
\end{table*}

\subsubsection{Number of flaring sources}\label{flaring}
For our third analysis, we need an objective way to estimate the number of flaring sources. In general, any source with activity index greater than one could be considered to be in a flaring state. However, especially in fainter objects, it is possible to obtain values greater than one simply due to noise in the data. 

To estimate a threshold for flaring sources that accounts for the typical uncertainties in the OVRO and Mets\"ahovi data, we simulate data of one window length with the same time sampling as the real data whose fluxes are varied  by the typical uncertainty in our data around a fixed constant value. The constant value is set to 0.15\,Jy for the OVRO and 1 Jy for the Mets\"ahovi samples to match a typical fainter target in each sample. This results in samples with no intrinsic variability, whose only variation is produced by the observing process.  We then calculate the mean flux density in this time window, and the "false A.I" is then the ratio between this mean flux density and the input constant flux (0.15\,Jy for OVRO and 1 Jy for Mets\"ahovi).
 We repeat this 100\,000 times to obtain a false detection rate. 

The limit for obtaining a higher activity index due to noise in 1\% of  the simulated cases is A.I.$>1.1$ in the OVRO data and A.I.$>1.16$ in the Mets\"ahovi data. A false detection rate of 0.01\% is obtained for A.I.$>1.29$ for the OVRO data and A.I.$>1.71$ for the Mets\"ahovi data. We use these limits to estimate the number of flaring sources in the samples, that is, the number of sources for which the A.I. in Table~\ref{table:ovro_ai} (OVRO) and Table~\ref{table:metsahovi} (Mets\"ahovi) is above these limits.

We then use these same limits for the random samples to obtain the probability of chance coincidence. The results from the comparison to random samples are summarized in Table~\ref{table:ai} for the mean activity index and flaring source analysis.

\begin{table*}
\caption{Chance coincidence for the activity index analysis}             
\label{table:ai}      
\centering          
\begin{tabular}{l l l l l l l l l l l l l l}     
\hline\hline      
Sample & N$_\mathrm{S}$  & $\Delta \psi$ & N$_\mathrm{A}$  & $<$A.I.$>$ & p & p & N$_\mathrm{f}$& p & p & N$_\mathrm{f}$ & p & p\\
 &  & (deg) & &  & pre & post &  (1\%) & pre & post  & (0.01\%) & pre & post\\
 (1) & (2) & (3) & (4) & (5) & (6) & (7) & (8) & (9) & (10) & (11) & (12) & (13) \\
\hline                    
OVRO all-AGN & 1795 & 0.9 & 18 & 1.15 & 0.003 & 0.007$(\times5)$ & 8 & 0.280 & 0.376$(\times5)$ & 6 & 0.019 & 0.027$(\times5)$ \\  
CGRaBS & 1157 & 0.9 & 17 &  1.20 & 0.010 & 0.023$(\times5)$ & 8 & 0.072 & 0.108$(\times5)$ & 6 & 0.005 & \textbf{0.006}$(\times5)$\\
OVRO-350mJy & 589 & 0.9 & 12 & 1.29 & 0.011 & 0.024$(\times5)$ & 7 & 0.013 & 0.018$(\times5)$ & 5 & 0.002& \textbf{0.003}$(\times5)$  \\
RFC-150mJy & 1156 & 0.9 & 16   & 1.21 & 0.005 & 0.014$(\times5)$ & 7 & 0.132 &0.193$(\times5)$ & 5 & 0.017 & 0.025$(\times5)$  \\
Mets\"ahovi & 183 & 0.5 &  7   & 1.26 & 0.096 & 0.158$(\times5)$ & 4 & 0.0012 & 0.0015$(\times5)$ & 1 & 0.057 & 0.058$(\times5)$\\
\hline                  
\end{tabular}
\tablefoot{Col (1) indicates the sample studied and col (2) gives the number of sources in that sample. Col (3) gives the optimal systematic uncertainty parameter $\Delta \psi$, which for each sample was found to be the same for all the A.I. analyses. Col (4) gives the number of associations found using this systematic uncertainty. The pre-trial and post-trial p-values are given in cols (6) and (7), respectively.  Col (8) gives the number of flaring sources in the sample when 1\% false-detection rate is used. The threshold is A.I. $> 1.1$ for the OVRO samples and A.I.$> 1.16$ for the Mets\"ahovi sample. Col (11) is the same for 0.01\% false-detection rate, which is A.I. $> 1.29$ for the OVRO samples and A.I.$> 1.71$ for the Mets\"ahovi sample.  The additional trial factor due to multiple samples is shown as a multiplicative factor in Cols (7), (10), and (13) (see Sect.~\ref{positions}). Values significant at the $2\sigma$ level when also this trial factor is accounted for are shown in bold for the samples considered to be statistically complete.}
\end{table*}

\section{Discussion}\label{discussion}
We looked at the connection between our radio observations of blazars and IceCube neutrino events by first associating the high-energy neutrino events listed in Table~\ref{table:neut} with our monitoring samples at OVRO and Mets\"ahovi.
In the following, we first discuss the number of associations in our samples, and the lack of potential counterparts in our samples based on other recent studies. We then discuss the implications of the small number of $\gamma$-ray-detected sources among our associations. Finally, we discuss our results on the mean flux density and activity index analysis, and their implications for future searches for the connection between neutrino events and blazars.

\subsection{Number of associated sources}

As can be seen from Tables \ref{table:ovro}, \ref{table:metsahovi}, and \ref{table:ovro_ai}, the number of neutrino events with associated blazars in our samples is in general small, namely from 6 to 15 ($11-27$\%), depending on the sample studied. Although we can expect that only $\sim 50\%$ of the 56 high-energy neutrinos in Table~\ref{table:neut} are of  astrophysical origin \citep{2019arXiv190711699I}, the number of associations is still much smaller than that. 

One potential explanation is the incompleteness of our monitoring samples. At OVRO, the monitoring began with the CGRaBS sample of 1157 objects, for which we find 15 associated events for both the activity index analysis (Table \ref{table:ovro_ai}) and the mean flux density analysis (Table \ref{table:ovro}). The number of associated sources is slightly higher than the number of associated events (17 and 18 for the activity index and mean flux density analysis, respectively) because there can be multiple sources within the error ellipses. Interestingly, including all AGNs monitored at OVRO does not change the associations drastically, although the number of objects in the sample increases to 1795. There are no additional neutrino events associated with OVRO sources and the increase in number of associated sources is simply due to the increase in the number of sources within the same error regions. 

This is interesting considering that most of the additional sources were included in the OVRO monitoring after they were detected by the Fermi-LAT, but 9 of the 20 associations listed in Table~\ref{table:ovro} have not been detected by Fermi-LAT (see also Sect.~\ref{sect:gamma}). This may imply that $\gamma$-ray activity does not play a major role in the neutrino association, and indeed recent papers on the association between $\gamma$-ray and neutrino emission have been inconclusive. \cite{krauss18} studied the $\gamma$-ray fluxes of all sources coincident with IceCube neutrino events at E$>100$\,TeV, and found no direct correlation between the $\gamma$-ray flux and expected neutrino flux. \cite{franckowiak20} also studied $\gamma$-ray detected sources associated with IceCube neutrino events at E$>100$\,TeV, and found that the results are consistent both with no correlation between the $\gamma$-ray and neutrino fluxes and with a linear correlation between the two. They also included the OVRO light curves of nine  associated sources, and found no direct link between the radio flux density and neutrino arrival times, although this was not quantified.

Table~\ref{table:sed} lists the optical classification, redshift, synchrotron peak frequency and luminosity, variability Doppler factor, and radio modulation index of the associated sources in the OVRO sample. The large majority of our 20 associations are FSRQs (14) with only 4 BL Lacs and 2 blazars of unknown type (BCU). This reflects the composition of our original CGRaBS sample, where 70\% of the sources are FSRQs and only 11\% are BL Lacs \citep{richards14}, and there is no contradiction with results of \cite{2019ICRC...36..916H} who found the strongest connection between LBLs and IBLs. Interestingly, one of our targets J2148+0657 has been identified in \cite{liodakis20} as a probable neutrino candidate based on a proton-synchrotron model. Although two other sources are within the error ellipse of IceCube (Fig.~\ref{ellipse}), confirming J2148+0657 as a neutrino emitter would have interesting implications for our understanding of jet energetics.

\begin{table*}
\caption{Properties of the associated OVRO sources} 
\label{table:sed}
\begin{tabular}{llllllll} 
\hline
\hline
J2000 Name & 4FGL Name & Opt. & $z$ & $\log(\nu_\mathrm{s})$ & $\log(\nu F\nu_\mathrm{s})$  & $\delta_\mathrm{var}$ & mod. \\
  & & class & & (Hz) & (erg cm$^{-2}$s$^{-1}$) & & index \\
  (1) & (2) & (3) & (4) & (5) & (6) & (7) & (8) \\
  \hline
  J0211+1051 & 4FGL J0211.2+1051 & BLL & 0.2 & 14.2 & $-$10.3 & 8.4$_{-2.1}^{+6.0}$ & 0.255\\
  J0336+3218 & 4FGL J0336.4+3224 & FSRQ & 1.259 & 13.2 & $-$11.0 & 5.5$_{-0.8}^{+8.8}$ & 0.275\\
  J0341+3352 & & FSRQ & 0.725 & 12.9 & $-$12.6& 5.1$_{-4.1}^{+3.3}$ & 0.227 \\ 
  J0502+1338 & 4FGL J0502.5+1340 & BLL & 0.35 & 13.0 & $-$11.7 & 15.8$_{-5.0}^{+5.4}$ & 0.301 \\ 
  J0509+0541 & 4FGL J0509.4+0542 & BLL & 0.3365 & 14.6 & $-$10.7 & 14.7$_{-5.1}^{+3.6}$ & 0.167\\
  J0630$-$1323 & & FSRQ & 1.021 & 13.6 & $-$12.1& 32.3$_{-30.1}^{+21.2}$ & 0.125\\
  J0631$-$1410 & & FSRQ & 1.017 & 12.6 &$-$12.2 & 0.6$_{-0.4}^{+2.7}$ & 0.100\\
  J1125+2610 & & FSRQ & 2.341 & 12.3 & $-$12.1& 56.8$_{-20.3}^{+27.0}$ & 0.140 \\ 
  J1256$-$0547 & 4FGL J1256.1$-$0547 & FSRQ & 0.5361 & 12.7 & $-$10.5 & 11.6$_{-1.1}^{+1.8}$ & 0.203\\
  J1310+3220 & 4FGL J1310.5+3221 & FSRQ & 0.997 & 12.7 & $-$11.2 & 26.4$_{-16.6}^{+13.4}$ & 0.244\\
  J1504+1029 & 4FGL J1504.4+1029 & FSRQ & 1.839 & 12.7 & $-$11.7 & 17.7$_{-6.1}^{+6.4}$ & 0.230\\
  J1743$-$0350 & 4FGL J1744.2$-$0353 & FSRQ & 1.057 & 12.7 & $-$11.2 & 6.4$_{-2.5}^{+15.3}$ & 0.056\\
  J1916$-$1519 & 4FGL J1916.7$-$1516 & BCU & $\dotsc$ & 12.8 & $-$12.0 & $\dotsc$ & 0.185\\
  J2148+0657 & 4FGL J2148.6+0652 & FSRQ & 0.99 & 12.5 & $-$10.7 & 1.2$_{-0.4}^{+0.9}$ & 0.201\\
  J2151+0709 & & FSRQ & 1.364 & 12.8 &$-$12.1 & 13.1$_{-4.7}^{+14.4}$ & 0.115\\
  J2151+0552 & & FSRQ & 0.74 & 12.4 &$-$12.2 & 18.0$_{-4.8}^{+4.3}$ & 0.030\\
  J2200+1030 & 4FGL J2200.3+1029 & BLL & 0.172 & 13.0 & $-$12.5 & 3.4$_{-2.8}^{+6.3}$ & 0.156\\
  J2200+0234 & & FSRQ & 1.323 & 13.1 &$-12.5$& 31.1$_{-5.8}^{+4.1}$ & 0.246\\
  J2203+1007 & & BCU\tablefootmark{b} & 0.172 & 12.6 &$-$12.5 & 7.6$_{-7.3}^{+14.6}$ & 0.156 \\
  J2238+0724 & & FSRQ & 1.011 & 12.9 &$-$12.0 & 16.8$_{-9.7}^{+7.3}$ & 0.347\\
  \hline\end{tabular}
  \tablefoot{Optical classifications in Col (3) are taken from the 4LAC catalog \citep{2019arXiv190510771T} for the sources included in there (ones with name given in Col (2)), and from the 5BZCat catalog \citep{2015Ap&SS.357...75M} for the non-LAT sources unless otherwise stated. BCU stands for blazars of unknown optical class. Col (4) gives the redshift of the source. Cols (5) and (6) give the synchrotron peak frequency and peak luminosity, respectively, taken from 4LAC catalog or determined by ourselves for the non-LAT sources. Variability Doppler factors, given in Col (7) are taken from \cite{2018ApJ...866..137L} or determined by ourselves using the same methods. Radio modulation index determined from OVRO data are given in Col (8), and are taken from \cite{liodakis17} or determined by ourselves.}
 \tablefoottext{a}{Classification taken from \cite{D_Abrusco_2014}}
 \tablefoottext{b}{Classification taken from \cite{Sowards_Emmerd_2005}}
  \end{table*}

As mentioned earlier, \cite{2020MNRAS.497..865G} found in their analysis an excess of ISP and HSP sources in the neutrino associations. We note that although the inclusion of the LAT-detected sources has increased the number of HSP sources in our monitoring sample, most of the sources are still LSP objects \citep{richards14}. 
We took the spectral energy distribution (SED) parameters of our $\gamma$-ray-detected associations from 4LAC \citep{2019arXiv190510771T} and modeled the non-$\gamma$-ray-detected ones ourselves using the Space Science Data Center SED builder{\footnote {\url{https://tools.ssdc.asi.it/}}}. These are also given in Table~\ref{table:sed}. Indeed a large majority of the sources are LSPs, and there are only two ISPs (TXS~0506+056 and J0211+1051) and no HSPs. When comparing our neutrino sample with the 3HSP sample \citep{2019A&A...632A..77C}, which was used as the parent sample in \cite{2020MNRAS.497..865G}, we find that there are eight possible neutrino associations with HSP sources that are missing from our radio monitoring sample. 

\cite{2020ApJ...894..101P} found that radio sources selected based on their compact 8\,GHz flux density and associated with neutrino events have on average higher flux densities than randomly selected objects, suggesting a connection between compact radio emission and neutrino events. As discussed earlier, their sample is based on the RFC catalog, and they use a flux density limit of 150\,mJy. Although we have included the cross-match of the RFC sample and the OVRO monitoring sample in our analysis, our current monitoring sample does not include all potential counterparts from RFC, and we lack several of the associations listed in \cite{2020ApJ...894..101P}. Cross-matching the full RFC catalog (above compact flux density of 150\,mJy at 8\,GHz and DEC$>-20^\circ$) of 2355 objects with our neutrino list using $\Delta \psi = 1.0^\circ$ gives 11 additional neutrino events, bringing the number of associated events up to 26, and the total number of associated objects to 32.

One important point to note is that the sample of \cite{2020ApJ...894..101P}, which is restricted to flux density greater than 150\,mJy, is missing most of the HSP sources that \cite{2020MNRAS.497..865G} include in their analysis, which are typically fainter in the radio band. Indeed, only 11 out of 84 HSPs in \cite{richards14} have intrinsic mean flux density above 150\,mJy at 15\,GHz. This shows that selecting the potential candidates from samples generated at single energy ranges may not reveal the complete picture. 

\subsection{Connection with $\gamma$-ray emission}\label{sect:gamma}

One significant feature on the list of associated OVRO sources is the large number of non-$\gamma$-ray-emitting sources, 9 out of 20. In general, some significant differences in properties of non-$\gamma$-ray-emitting and $\gamma$-ray-emitting blazars have been found and these could be of relevance also to the expected neutrino fluxes, and therefore we discuss these properties in the following. 

\cite{savolainen10} found that $\gamma$-ray-detected blazars had in general higher Doppler factors than non-$\gamma$-ray-detected ones. Later studies \citep{richards14, 2015ApJ...810L...9L} confirmed that the non-$\gamma$-ray blazars had smaller 15\,GHz modulation indices (defined as the standard deviation of the flux density variations over the mean flux density) and lower apparent jet speeds, both suggesting lower Doppler boosting. Therefore, we investigated modulation indices and variability Doppler factors of the $\gamma$-ray and non-$\gamma$-ray-detected sources in our sample.

 For some of the sources in our sample, the radio modulation indices using the OVRO 15\,GHz data were available in \cite{liodakis17}, and we used the same method to estimate them for the remaining sources. The variability Doppler factors were also estimated using the variability timescales and amplitudes from the 15\,GHz OVRO light curves with the method\footnote{We note that the variability Doppler factors depend only on the assumed emission region geometry and intrinsic brightness temperature, and are therefore less model dependent than those derived from SED fitting \citep[see e.g.,][who compared different methods to derive Doppler factors]{LV99}.} described in \cite{2018ApJ...866..137L}, which already included these for a subset of our sample. Again, for the remaining sources these were estimated in the same manner.

The modulation indices and  variability Doppler factors for the sources are given in Table~\ref{table:sed}. We performed a Kolmogorov-Smirnov test and found that there are no significant differences between the Doppler factors (p-value 0.49) or 15\,GHz modulation indices (p-value 0.19) of the non-$\gamma$-ray and $\gamma$-ray sources in our sample.
For the non-$\gamma$-ray sources in our sample, the mean modulation index is 0.165 and the mean variability Doppler factor is 20.2, while for the $\gamma$-ray sources these values are 0.207 and 11.1. 
Comparing these to average modulation indices for $\gamma$-ray blazars, namely $0.175^{+0.012}_{-0.011}$, and for non-$\gamma$-ray blazars, namely $0.099^{+0.004}_{-0.003}$ \citep{richards14}, and to median Doppler factors for the $\gamma$-ray-detected ($\delta_\mathrm{var}=14$) and non-detected  ($\delta_\mathrm{var}=8$) from \cite{2018ApJ...866..137L} demonstrates that the non-$\gamma$-ray-detected sources in our sample are actually as variable and as Doppler-boosted as typical $\gamma$-ray blazars, and from that point of view they are not less likely to be neutrino sources than the average $\gamma$-ray blazar.

As discussed by for example \cite{richards14} and \cite{2015ApJ...810L...9L} the non-$\gamma$-ray-detected blazars also have lower synchrotron peak frequencies than the $\gamma$-ray-detected blazars. As our sample mainly consists of LSP sources, comparing the SED classes of our $\gamma$-ray-detected and non-$\gamma$-ray-detected sources is not of interest. We note however that our non-$\gamma$-ray-detected sources all have low $\nu_\mathrm{S}$ values. The range is $\log(\nu_\mathrm{S})=12.3-13.6$, with only two sources with $\log(\nu_\mathrm{S})\geq13$, while for example \cite{keenan20} found the average for LSP and FSRQ sources to be $\sim$13.4. This of course largely explains why they are not detected in  $\gamma$-rays; with typically $\sim10^7$ between the peaks \citep{fossati98}, the second peak in the SED is not in the {\it Fermi}-LAT energy range. The nature of these sources is little studied as most of the "blazar sequence" studies are concentrated on $\gamma$-ray-detected blazars \citep[e.g.,][]{ghisellini17}. We note however that we also estimated the $\nu F_\nu$ of these sources from the SED fits (see Table
~\ref{table:sed}). The non-$\gamma$-ray-detected sources in our sample are fainter than the $\gamma$-ray-detected ones, but still have
$L_{\mathrm{peak}}=10^{44.9}-10^{46.8}$ erg$/$s, and so these sources are not low-luminosity sources in general. 

The actual conditions and predictions of the neutrino rates from the sources discussed in our paper would require detailed modeling and this is beyond the scope of the current paper. We simply note that there is no obvious reason to exclude any of the sources as possible neutrino emitters. This is also in line with the discussion in \cite{plavin20b}, who did not find reasons to disfavor the non-$\gamma$-ray-detected AGNs as neutrino sources in terms of target photon fields or the accelerated protons.

\subsection{Mean flux density and variability of the radio emission}
As discussed earlier, although we have used several subsets of the OVRO monitoring sample to look at the connection between radio emission and IceCube neutrino events, the number of associated sources is very similar in the different samples. Therefore, it is not surprising that the results for these different samples are in general consistent, as can be seen from Tables \ref{table:meanflux} and \ref{table:ai}.  We note again that the OVRO all-AGN and Mets\"ahovi samples are not statistically complete, and so those results should be treated with caution, and we do not account for them when assessing the significance of our results.

Unlike \cite{2020ApJ...894..101P}, we do not see a highly significant difference between the mean flux density of the associated sources compared to the other sources in our samples (Table~\ref{table:meanflux}). 
However, when we look at the radio variability of these sources, our results are suggestive of a connection between the neutrino events and radio flares (Table~\ref{table:ai}). Although the results for the mean activity index are not significant when the sample trial factor is accounted for (except for the OVRO all-AGN sample, which we do not consider statistically complete), we do see a significant effect when we look at the number of flaring sources for both the CGRaBS and OVRO-350mJy samples. For example, the chance probability of finding five strongly flaring blazars (A.I.$>1.29$ with 0.01\% false-detection limit for flaring sources) of the CGRaBS sample coincident with neutrino arrival times is only 3\% (post-trial with sample correction also applied). These sources are not necessarily the brightest sources in the sample, and therefore they are not the same objects that \cite{2020ApJ...894..101P} indicated as the most likely neutrino-emitting sources based on their mean flux density.

For the Mets\"ahovi sample, we also do not find a significant effect when we look at the average of the activity indices, but the chance probability of finding four flaring blazars (A.I.$>1.16$) coincident with neutrino arrival times is only 0.75\% (post-trial). This value is obtained when using the 1\% false-detection limit for defining the threshold for flaring blazars. If we use the 0.01\% limit of A.I.$>1.71$, there is only one source in our sample above that limit, also making the chance probability much higher.
  We note that these values should only be considered illustrative due to the incompleteness of the parent sample.

We note that the number of strongly flaring blazars in both OVRO and Mets\"ahovi samples is very small and the significance of the analysis is somewhat sensitive to the false-detection limits used because inclusion or exclusion of individual sources changes the results. Moreover, the number of strongly flaring blazars is very small compared to the number of IceCube neutrino events even if 50\% of them are not astrophysical. Therefore, it is clear that not all neutrino events happen when a co-spatial blazar shows strong radio activity. However, when we see strong radio activity in a blazar and a neutrino event at the same time, it is unlikely to be a random coincidence. This is consistent with the association of PKS~B1424-418 with the high-energy neutrino IC35 by \cite{2016NatPh..12..807K} who found the source to be flaring in both $\gamma$-ray and radio bands at the time of the neutrino arrival.  Based on a sample of four possible neutrino associations and their OVRO light curves (including J0506+056 and J1504+1029 from our sample), \cite{weber20} also note that the neutrino arrival times seem to coincide with flaring in the radio band.

Finally, although the activity index is a simple way to estimate the flaring state of a blazar, it is not ideal when using light curves that are several years long and well sampled. For example, in the source J1743$-$0350, the coincident neutrino arrives during the rising part of a large outburst (Fig.~\ref{Figpage2}) but because the light curve includes other flares of similar strength, the overall mean flux density of the light curve is high, resulting in a smaller activity index. Another similar case is J1125+2610 (Fig.~\ref{Figpage1}). Even in the case of TXS~0506+056 (J0509+0541, Fig.~\ref{Figpage1}), the activity index does not indicate flaring because the radio outburst continues after the neutrino arrival. 

The activity index and mean flux density in the radio bands are also not appropriate metrics for associating the HSP sources because in addition to being faint in the radio frequencies, the HSP sources are also less variable \citep{richards14}. Other ways to combine the radio and $\gamma$-ray catalogs should be investigated, in addition to more sophisticated methods for comparing radio flares to neutrino arrival times, but these are beyond the scope of this paper.

\subsection{Comparison between OVRO and Mets\"ahovi associations}
As discussed earlier, \cite{2020ApJ...894..101P} found the strongest temporal connection between the RATAN-600 highest frequency data at 22\,GHz and the neutrino arrival times. Although the Mets\"ahovi sample is statistically incomplete, it is interesting to compare the light curves and activity index values of the individual associations in the two different frequencies. All but one of the Mets\"ahovi associations (1308+328, which was eliminated from the OVRO sample, see Sect. \ref{ovro}) are also included in the OVRO associations. In general, the light curves at 37\,GHz show much faster variations than at 15\,GHz \citep[e.g.,][]{hovatta07}, which is typically attributed to higher opacity in the jets at lower frequencies.

This is also seen when comparing the light curves of the associated sources. As discussed in the previous section, there are four associations in the Mets\"ahovi sample that show strong radio flares at the time of the neutrino arrival. One of these is J1743$-$0350, which did not meet the formal criteria of strong flares in the OVRO data as discussed above. In the Mets\"ahovi light curve (Fig.~\ref{Figmhovi}), it can be seen that the neutrino arrival time coincides with an extremely fast increase of the radio flux density at 37\,GHz. At 15\,GHz, the substructures of the flares are not as clearly visible, but there the neutrino arrival time also coincides with a fast increase of flux density.
The other three flaring sources in the Mets\"ahovi sample are also identified as strongly flaring sources in the OVRO light curves, and overall the activity index values between the two data sets are similar. 

We also note that while there is a large fraction of non-$\gamma$-ray-detected sources in the OVRO associations (see Sect.~\ref{sect:gamma}), all the Mets\"ahovi associations are $\gamma$-ray detected. This is mainly due to the smaller number of objects monitored regularly at Mets\"ahovi, and the higher flux density limit used in our analysis. As can be seen in Table~\ref{table:ovro}, all the non-$\gamma$-ray-detected sources have $\bar S_{15\,GHz}<0.9\,Jy$.

\section{Conclusions}\label{conclusions}
We studied the possible connection between high-energy neutrinos and blazars by comparing our radio monitoring catalogs from OVRO and Mets\"ahovi with positions of IceCube neutrino events. The number of associations we find is in general small. Only $11-27$\% of the 56 neutrino events are found to be associated with blazars, depending on the radio sample used. The small number of associations can be largely explained by incomplete parent samples of radio sources used in this study and is not in contradiction with the recent studies of \cite{2020ApJ...894..101P} and \cite{plavin20b} associating the observed astrophysical neutrino flux to radio bright AGNs.

Our associated sources are largely LSP FSRQs and a large fraction of the OVRO associations are not  detected in $\gamma$-rays (see Sect.~\ref{sect:gamma}). This reflects the composition of the OVRO monitoring sample, which largely consists of FSRQs and non-$\gamma$-ray-detected sources, meaning that our study is not suitable for identifying the most likely class of sources associated with neutrinos.

We find that the luminosity and radio variability characteristics ---including variability Doppler boosting factors--- of the non-$\gamma$-ray-detected associations in our sample are indistinguishable from the $\gamma$-ray-detected associations. Because these sources are mostly  luminous FSRQs, 
they are also potential neutrino emitters, which are often ignored in studies concentrating only on $\gamma$-ray-detected sources.

When studying the radio emission and variability of the associated sources, we find no difference in the mean flux density of the associated sources compared to random control populations. However, when radio variability amplitudes are studied, we find a connection between the largest flares and neutrino arrival times, and conclude that when we see strong radio activity in a blazar and a neutrino event at the same time, it is unlikely to be a random coincidence.

This is in line with the results of recent studies of the connection between blazar $\gamma$-ray flares and the production of detectable neutrino fluxes. It is evident that the fluence of even the brightest keV-GeV flares is not sufficient to result in a likely detection of neutrinos and the only way to increase this likelihood is to increase the duration of the flare \citep{2019MNRAS.489.4347O, 2020arXiv200900125K}. Strong radio flares have a typical duration of 1-2 yr (see Section 3.3.) and are often accompanied with long-duration flaring at higher energies. Therefore, they are natural candidates to be associated with detectable neutrino emission and we will study this connection in more detail in future works.

\section*{Acknowledgements}
We thank A. Plavin, Y. Kovalev, M. Petropoulou, A. Franckowiak, M. Kadler, P. Hakala, and V. Pavlidou for useful discussions, and K. Wiik for help in setting up the computations. The OVRO 40 m program was supported by NASA grants NNG06GG1G, NNX08AW31G, NNX11A043G, and NNX13AQ89G from 2006 to 2016 and NSF grants AST-0808050, and AST-1109911 from 2008 to 2014, along with private funding from Caltech and the MPIfR.
T. H. was supported by the Academy of Finland projects 317383, 320085, and 322535.
E. L. was supported by the Academy of Finland projects 317636 and 320045.
S.K. acknowledges support from the European Research Council (ERC) under the European Unions Horizon 2020 research and innovation programme under grant agreement No.~771282. 
W.M. acknowledges support from ANID projects Basal AFB-170002 and PAI79160080. R.R. acknowledges support from ANID Basal AFB-170002, and ANID-FONDECYT grant 1181620. Handling of catalogs in the paper was partially done using TOPCAT \citep{topcat}. The computer resources of the Finnish IT Center for Science (CSC) and the FGCI project (Finland) are acknowledged.

\bibliographystyle{aa} 
\bibliography{neutrinos.bib} 

\begin{appendix} 
\section{OVRO 15\,GHz and Mets\"ahovi 37\,GHz light curves of the associated sources}
   \begin{figure*}
   \centering
   \includegraphics[width=\hsize]{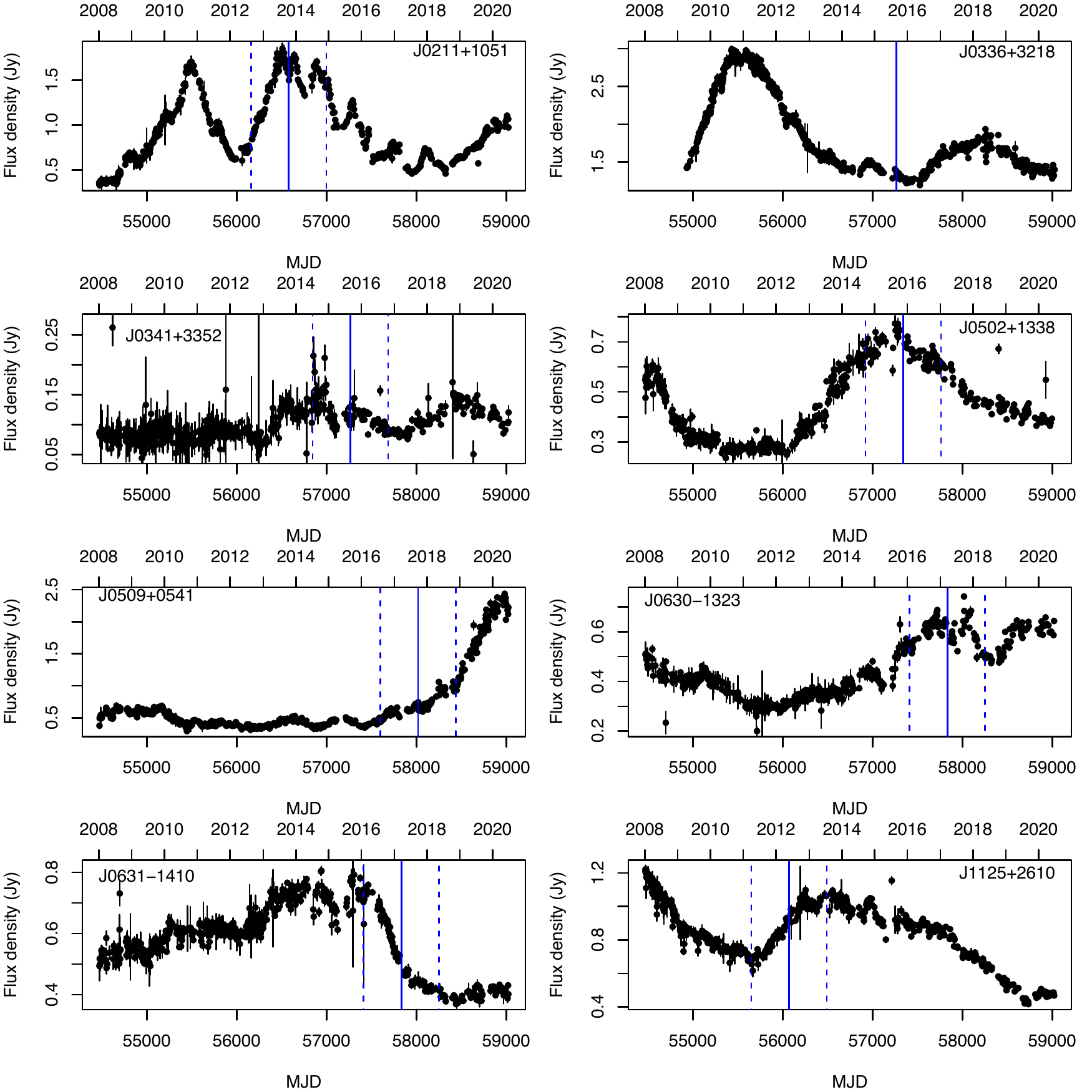}
      \caption{OVRO light curves of the associated sources. The neutrino arrival time is shown with a blue solid line, while the blue dashed lines indicate the window of 2.3 yr around the neutrino event for the sources included in the activity index analysis.}.
         \label{Figpage1}
   \end{figure*}

   \begin{figure*}
   \centering
   \includegraphics[width=\hsize]{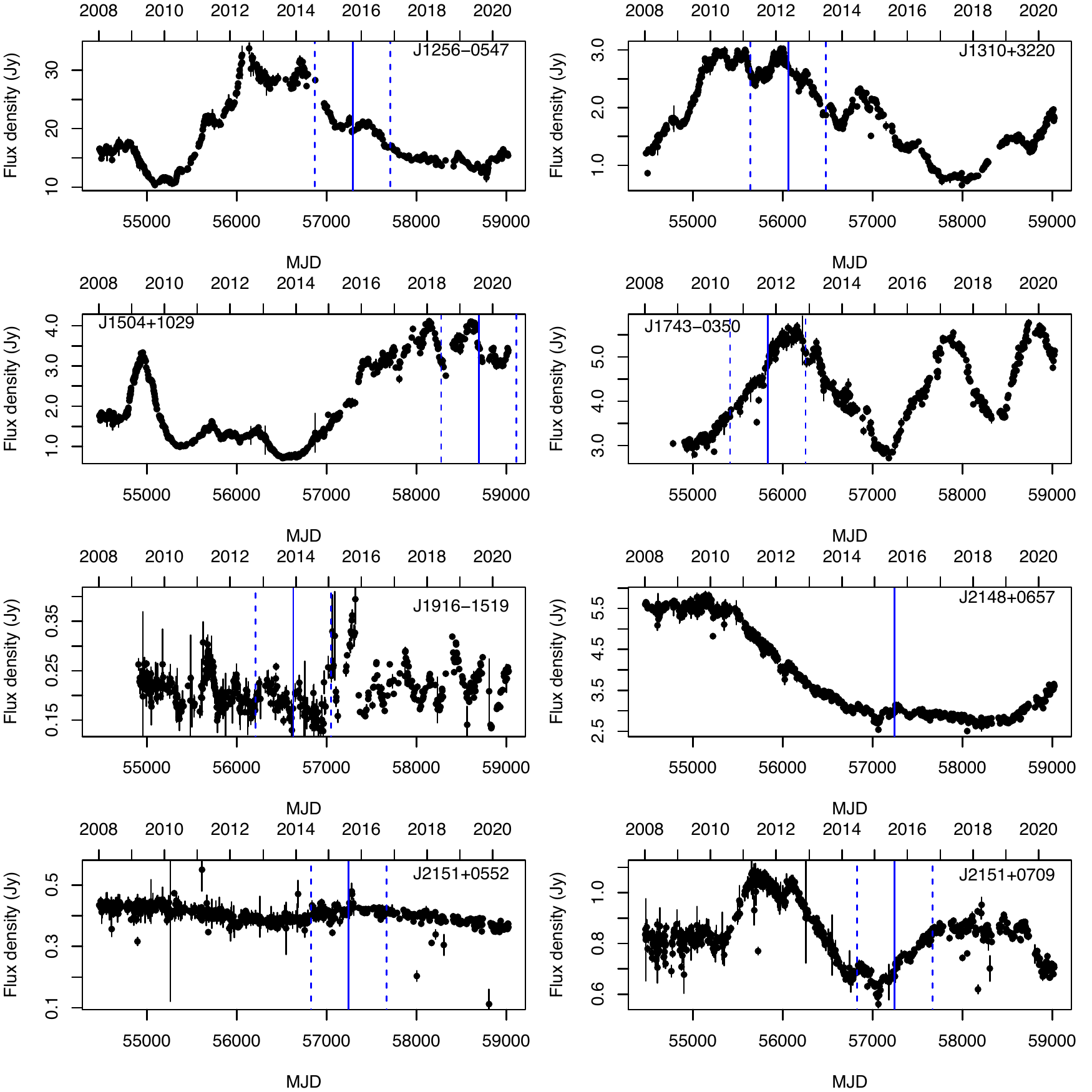}
      \caption{OVRO light curves of the associated sources. The neutrino arrival time is shown with a blue solid line, while the blue dashed lines indicate the window of 2.3 yr around the neutrino event for the sources included in the activity index analysis.}
         \label{Figpage2}
   \end{figure*}

   \begin{figure*}
   \centering
   \includegraphics[width=\hsize]{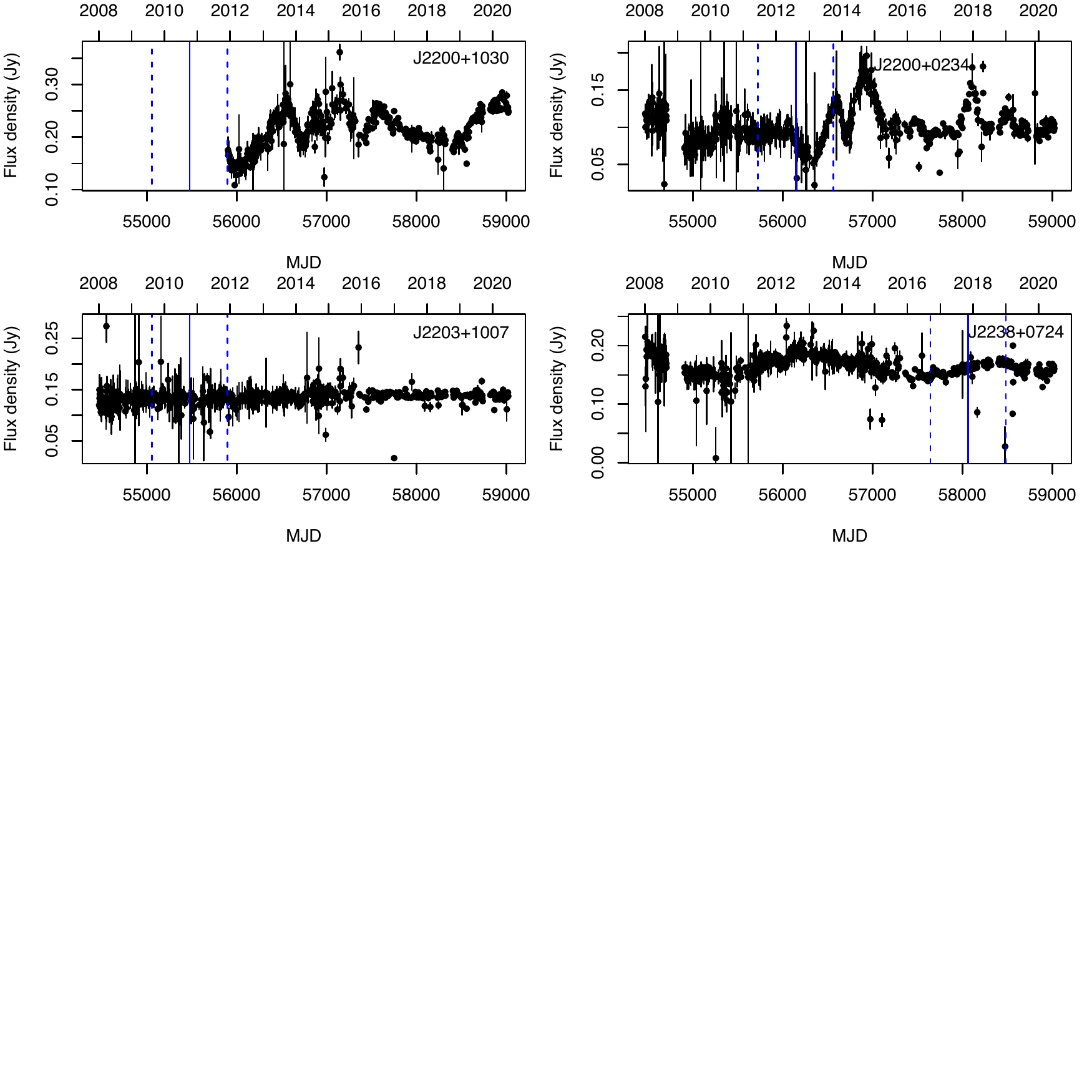}
      \caption{OVRO light curves of the associated sources. The neutrino arrival time is shown with a blue solid line, while the blue dashed lines indicate the window of 2.3 yr around the neutrino event for the sources included in the activity index analysis.}
         \label{Figpage3}
   \end{figure*}

 \begin{figure*}
   \centering
   \includegraphics[width=\hsize]{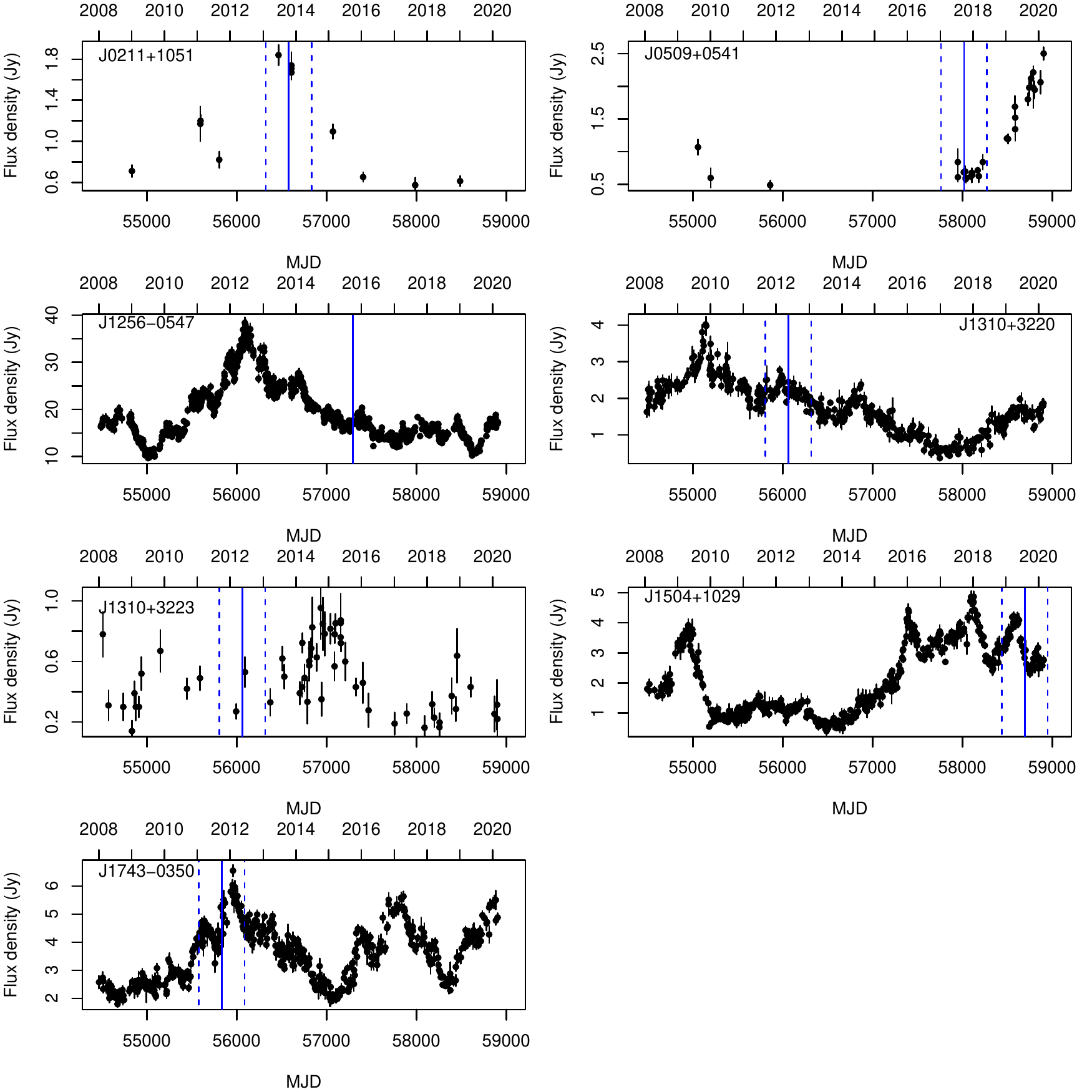}
      \caption{Mets\"ahovi light curves of the associated sources. The neutrino arrival time is shown with a blue solid line, while the blue dashed lines indicate the window of 1.4 yr around the neutrino event for the sources included in the activity index analysis.}
         \label{Figmhovi}
   \end{figure*}

\end{appendix}
\end{document}